\documentclass[draft,pdflatex]{agujournal2019}
\usepackage{url} 
\usepackage{lineno}
\usepackage[inline]{trackchanges} 
\usepackage{soul}
\usepackage{placeins}
\usepackage[outdir=./]{epstopdf}
\draftfalse

\journalname{JGR: Earth Surface}

\begin{document}

%
%


\title{Basal force fluctuations and granular rheology: Linking macroscopic descriptions of granular flows to bed forces with implications for monitoring signals}

%
%




\authors{P. Zrelak\affil{1}, E. C. P. Breard\affil{1}$^,$\affil{2}\thanks{Current address}, J. Dufek\affil{1}}

\affiliation{1}{Department of Earth Sciences, University of Oregon, Eugene, Oregon, U.S.A}
\affiliation{2}{School of Geosciences, The University of Edinburgh, Edinburgh, Scotland, U.K}





\correspondingauthor{P.J. Zrelak}{pzrelak@uoregon.edu}




\begin{keypoints}
\item Discrete element simulations of granular flows in plane-shear and inclined-slope flow configurations reveal bed contact force distributions
\item Basal forces record changes in flow regime correlating with non-dimensional shear rate (inertial number) independent of flow configuration
\item Basal forces record a transitional regime from liquid-like to gas-like granular dynamics in which memory stored via elastic contacts are erased
\end{keypoints}

%
%

%
%


\begin{abstract}
Granular flows are ubiquitous in nature with single flows traversing a wide range of dynamic conditions from initiation to deposition. Many of these flows are responsible for significant hazards and can generate remotely detectable seismic signals. These signals provide a potential for real-time flow measurements from a safe distance. To fully realize the benefit of seismic measurements, basal-granular forces must be linked to macroscopic internal flow dynamics across a wide range of flow conditions. We utilize discrete element simulations to observe dry and submerged granular flows under plane-shear and inclined-flow configurations, relating bulk kinematics to basal-force distributions. We find that the power and frequency of force fluctuations scale with non-dimensional shear rate ($I$). This scaling tracks three pre-established regimes that are described by $\mu(I)$ rheology: (1) an intermittent particle rearrangement regime, where basal forces are dominated by low frequencies; (2) an intermediate regime where basal forces start to increase in frequency while showing correlations in space; and (3) a fully collisional regime where the signal is nearly flat up to a cutoff frequency. We further identify a newly defined fourth regime that marks a `phase change' from the intermediate to collisional regime where increases in basal force fluctuations with increasing shear rates stalls as the granular bed dilates, partially destroying the contact network. This effort suggests that basal forces can be used to interpret complex granular processes in geophysical flows.
\end{abstract}

\section*{Plain Language Summary} Debris flows, pyroclastic flows, and debris avalanches are natural granular flows that have the ability to generate seismic signals. These signals can be recorded from safe distances away, providing a record of how these flows interact with the surface. Nevertheless, it is difficult to interpret these signals in terms of flow conditions. To better understand how to relate the forces exerted by these flows back to flow conditions, we perform simulations of granular flows, recording the forces exerted on a simulated substrate. Simulated flow conditions resemble those that correspond to the beginning of an event, when the flow is slow moving, to conditions that are like that of a fast moving well developed flow. We find that the forces exerted on the ground have the ability to tell researchers important flow information, such as how stresses are transmitted internally in the flow. Moreover, these basal forces record when the collection of grains changes from a liquid-like to a gas-like flow, where individual grains are no longer in prolonged contact, but begin to collide and bounce off one another, like gas-molecules. These results show that complex information can be encoded in the interactions between an overriding flow and it's substrate.

%
%

%


%
%
%
%
\section{Introduction}\label{sec1} \FloatBarrier
Flowing granular mixtures are prevalent in nature. Geophysical mass flows experience a range of granular interactions due to spatio-temporal variations in particle concentration, particle-fluid-particle interactions, grain size distributions, and shear rates. Debris flows, pyroclastic density currents, and debris avalanches showcase complex phenomenological behavior that is intrinsic to granular materials. In this view, debris flows are fluid-particle mixtures that can span from dilute, near hyperconcentrated flow conditions, to dense flows able to support large clasts \cite{Iverson1997}. Pyroclastic density currents, gravity flows of volcanic particles and gas, display enhanced mobility that is attributed to the liquid-like nature of gas fluidized granular flows and the development of pore pressure (e.g., \citeA{Lube2019, Breard2023}). Each of these disparate phenomena can span from dry conditions, where particle inertia allows for drag effects to be neglected, to fluid immersed conditions, where fluid drag and pore pressure gradients become important \cite{Delannay2017, Lube2020}. Moreover, these geophysical flows can produce remotely detectable seismic signals (e.g., \citeA{Allstadt2018}). It is our goal to aid in developing a framework relating basal forces that drive seismicity to bulk rheology. In this effort, we will report on data derived from granular flow numerical simulations utilizing discrete element methods where individual particle motion is explicitly resolved in the presence, and absence, of an interstitial continuous fluid. We will use these simulations to observe flow rheology under disparate shear states which can be expressed via non-dimensional parameters, allowing us to compare our scaled simulations to large scale geophysical mass flows. During these simulations we will be recording basal forces exerted on a simulated force plate, allowing us to relate bulk flow rheology to measured forcings.

Several processes are involved in the generation and transmission of seismic waves from source to receiver. A recorded seismic signal e.g., ground velocity $\dot{u}(t)$, measured at some distance from the flow, is a result of a convolution of three functions:
\begin{equation} \label{conv}
    \dot{u}(t) = F(t) * g(t) * r(t)
\end{equation}
with $F(t)$ being the time-varying forces exerted on the substrate (i.e., source function), $g(t)$ being a transfer function --- known as the green's function --- representing path effects that attenuate and disperse forcing energy, and $r(t)$ being the instrument impulse response (e.g., \citeA{Stein2003}; Fig. \ref{SETUP}). Much effort has been invested in utilizing geophysical signals to inform scientists and hazard coordinators, with the aim to improve detection and monitoring. For example, time and spectral characteristics of seismic signals have been used to train machine learning algorithms to detect debris flow events (e.g., \citeA{Chmiel2021}); seismic envelopes have been correlated to event volume (e.g., \citeA{Levy2015}) and potential energy loss (e.g., \citeA{Farin2018}); and the ratio of horizontal and vertical ground displacements have been used to posit turbulent flow conditions \cite{Cole2009,Walsh2020}. Nevertheless, the link between internal dynamics and the source function $F(t)$ is a non-trivial one. \citeA{Allstadt2020} report data from controlled, large scale ($8-10\ m^3$), debris-flow experiments and calculated empirical green's functions, constraining the path effects term $g(t)$. This allows the authors to invert basal tractions from seismic signals recorded by near-channel receivers. These inverted forces compare well with in-channel force plate measurements. However, the authors were limited in their ability to use these inverted signals to infer detailed flow information due to the poorly constrained relationship between macroscopic flow dynamics and boundary forcing. 

Measuring boundary forces, thus constraining the source function $F(t)$, is important to understand not only how seismic forces are generated, but how flows affect and are affected by their environment. \citeA{McCoy2013} measure basal forces in natural debris flows, relating these basal forces to channel incision. These authors leverage the high rate of debris flow events per year at their study site (Chalk Cliffs, CO), instrumenting the channel with a variety of sensors, creating a natural laboratory. Nevertheless, collecting data from real-world flows is difficult, providing the need for controlled experiments. Large-scale debris-flow experiments have been conducted at a controlled test site in Blue River, Oregon, USA since 1994 \cite{Iverson20102}. Here, a wealth of data is collected and disparate sensor arrays supply multiple data streams to analyze a flow. These experiments not only provided data for \citeA{Allstadt2020} but also for the development of depth-averaged models \cite{Iverson2014,George2014}, and, in general, led to a more refined understanding of the physics of debris flows \cite{Iverson1997}. Large-scale experiments are needed to reliably recreate near-scale geophysical flows.  Nevertheless, as experimental scale increases, so does uncertainty. For example, near-scale experiments utilizing natural particles (e.g., sand and gravel) may have complex shape and size distribution, which may lead to unforeseen phenomenon. Further, steady states are difficult to maintain when conducting large experiments. These uncertainties, paired with the resources required for conducting such experiments, highlights the need for smaller, scaled experiments. Researchers have utilized scaled experiments in various configurations to observe the probability distribution of boundary forces and the dynamical nature of these forces as they approach the jamming transition \cite{Longhi2002,Corwin2005,Gardel2009,Langroudi2010}. The development of force chains in granular flows has been shown to create bed-force localization that can greatly exceed mean forces \cite{Estep2012}. Other authors have looked at forces generated during individual particle-impact events, deriving scaling laws relating the mass and speed of the particle to energy transferred to the boundaries \cite{Farin2015}. Recent work has performed monodisperse inclined-slope flow experiments, recording basal forces at high frequencies, comparing these data to modeled data from stochastic impact models \cite{Arran2021}. Importantly, these authors find that the normalized squared basal-force fluctuation, in the intermediate and high frequency ranges associated with uncorrelated particle interactions, is related to the bulk scaled shear rate.

In the limit where grain-inertia dominates flow behaviour (e.g., unsaturated debris avalanches, the snouts of particle concentrated pyroclastic flows, and the coarse front of debris flows), substrate forcing, like bulk flow dynamics, is expected to be controlled by particle-particle interaction  \cite{Forterre2008}. Previous efforts show that sheared granular media span three distinct regimes that are related to the shear rate: (1) stick-slip unsteady motion, (2) continuous deformation of the granular bed, (3) high energy collisional flow, with the nature of particle velocity fluctuations being controlled by these regimes \cite{daCruz2005}. Velocity fluctuations are what give rise to a quantity referred to as granular temperature \cite{Lun1984}. Much like the thermal temperature of a gas, this quantity measures fluctuations in kinetic energy of individual particles in a system, generating a variance in the particle velocity distribution. While often treated as isotropic, heterogeneties in granular flows may give rise to local anisotropy in granular temperature, i.e. velocity fluctuations in different directions may vary in magnitude. As opposed to molecular gasses, granular systems are by definition in a perpetual state of dis-equilibrium due to energy being dissipated through inelastic particle-particle and particle-domain interactions \cite{Goldhirsch2008}. \citeA{Trulsson2012} show that the dissipation mechanism in dense granular suspensions is a function of confining pressure $P_s$ scaled by the interstitial fluid's viscosity $\eta_f$: $\sqrt{\rho_s P_s}d/\eta_f = I/\sqrt{I_v}$, where $\rho_s$ is the solid particle density, $I$ is the inertial number from \citeA{daCruz2005}, and $I_v$ is  the viscous inertial number from \citeA{Boyer2011} (Eqs. (\ref{I}) and (\ref{viscous_inert}), respectively). As this scaled confining pressure (i.e., $I/\sqrt{I_v}$) approaches 10, contact forces increase rapidly to a limit where dissipation is controlled solely by particle contacts, and the flow behaves as a pure granular flow. In other terms, the dissipation mechanism in fluid immersed flows is dictated by the competition between confining pressure (this can be referred to as the solids pressure, or the pressure that the granular material exerts upon itself), and viscous effects. Further, scaling laws relating wall friction to slip velocity scaled by velocity fluctuations provides evidence that fluctuations are key to describing dense flows near wall boundaries \cite{Artoni2015}. This suggests that flow information may be extracted from force fluctuations, or through the measurement of energy dissipation,  near the flow-boundary. 

While much progress has been made on end-member conditions, detailed information of bed-force distributions across a range on inertial numbers is still lacking. Here we perform a series of three-dimensional numerical simulations utilizing the discrete element method (DEM) to examine granular contributions to wall forcing, thus observing the source-time function $F(t)$ in Eq. \ref{conv} as a function of flow state. The discrete element approach resolves the interactions of individual particles with the bottom boundary, enabling a high resolution record of bed forces across a range of flows at different flow states. With DEM, we can track the forces associated with each individual particle in the system, rather than estimating granular stresses through, for example, theoretical relations or phenomenological approximations between flow properties and particle forcings that may rely on simplifying assumptions. The goal of this study is to relate macroscopic descriptions of granular flows to basal forcing under ideal conditions, but over a wide range of inertial numbers. We will first briefly describe our methods, pointing the reader to preceding studies that advanced these efforts. Then, in Section \ref{Results}, we will detail bed-averaged flow conditions, afterwards discussing basal forces recorded under these conditions in Section \ref{WALL_FORCES}. We will close with a discussion relating the two observations culminating in a phase space spanned by our data in Section \ref{PHASE}, and examine the implications of these efforts as we scale towards real-world flows in Section \ref{REASCALE}.
 \FloatBarrier
\section{Methods}\label{sec2}

\subsection{MFIX-DEM}\label{MFIX} 
Our three-dimensional numerical simulations are conducted using the Department of Energy MFiX-DEM solver. Detailed information on MFiX-DEM is discussed in \cite{Garg2012}. MFiX-DEM solves the ordinary differential equations governing the motion of the $i^{th}$ particle\\
\begin{equation} \label{NEWTS}
m_i \frac{d\mathbf{V}_i}{dt} = m_i\mathbf{g} + \mathbf{F}_{d}(t) + \mathbf{F}_c(t)   
\end{equation}
\noindent where $\mathbf{V}_i$ is the velocity of the particle, $\mathbf{g}$ is the gravity vector, $\mathbf{F}_c(t)$ is the net contact forces associated with contacting particles, $m_i$ is the mass of the $i^{th}$ particle, $\mathbf{F}_{d}(t)$ is the drag forces acting on the $i^{th}$ particle, and $t$ is time. We perform our simulations under steady state conditions. Consequently, the added mass and history terms of the generalized Basset-Boussinesq-Oseen equation, arising due to unsteady fluid flow relative to the solid particles, are neglected \cite{MaxeyandRiley}. We use the soft-sphere approach to particle contacts, modeling collisions and particle overlap as interactions between systems composed of springs and dashpots. In this way, the spring supplies the energy for rebound, gaining potential energy during collision and releasing post-collision; the dashpot simulates dissipation due to inelasticity. Thus, the net contact force in Eq. (\ref{NEWTS}), in the normal direction, becomes a sum of the elastic ($F^n_{k_n}$) and viscous ($F^n_\eta$) components
\begin{equation} \label{FC}
F^n_c = F^n_{k_n}+F^n_\eta = -(k_n \delta_n + \eta_n |V_{n}^{i} - V_{n}^{j}| )
\end{equation}
where scripts \textit{n} in the above and subsequent equations represent parameters in the normal direction. The net contact force is used to visualize force chain magnitude, where the magnitude is the norm of the contact forces. The midpoint of a single `link' in a force chain is the contact point between two particles, and the length of this link is the distance between the centers of the two particles. Interconnected links of this kind constitute a force chain. 
The elastic contribution in Eq. (\ref{FC}) is controlled by the elastic coefficient $k_n$ and the particle overlap in the normal direction, $\delta_n$. The viscous contribution is a function of the dampening coefficient $\eta_n$ and the relative normal velocities $V_{n}$ of the $i^{th}$ and $j^{th}$ contacting particles. The dampening coefficient is dictated by the elastic coefficient, normal coefficient of restitution $e_n$, and the effective mass $m_{ef}$ between the two colliding particles
\begin{equation} \label{DAMP}
\eta_n = \frac{2\sqrt{m_{ef}k_n}|ln\ e_{n}|}{\sqrt{\pi + ln^2\ e_n}}
\end{equation}
\noindent Tangential elastic and viscous coefficients are related to their normal counterparts by a simple scale factor (i.e. $k_t = \frac{2}{5}k_n$ and $\eta_t = \frac{2}{7}\eta_n$). The minimum collision time between two particles is dictated by the elastic and viscous coefficients:\\
\begin{equation} \label{TCOL}
    t_n^{col} = \pi \left(\frac{k_n}{m_{ef}} - \frac{\eta_n^2}{4m_{ef}^2} \right)^{-1/2}
\end{equation}
\noindent Timesteps are set to $0.2t_n^{col}$ to resolve collisional interactions.  

\indent In the case where particles are submerged in a fluid, continuum mechanics are introduced and the fluid is governed by continuity and momentum conservation detailed in \cite{Syamlal1993}:
\begin{equation} \label{FLUID}
    \frac{D}{Dt}(\epsilon_f \rho_f \mathbf{u}_f) = \mathbf{\nabla} \cdot \overline{\overline{S}}_f + \epsilon_f \rho_f \mathbf{g} - \mathbf{I}_{f}
\end{equation}
\noindent Where $D/Dt$ is the material time derivative, $\epsilon_f$ is the fluid volume fraction, $\rho_f$ is the fluid density, $\mathbf{u}_f$ is the fluid velocity, and $\overline{\overline{S}}_f$ is the fluid phase stress tensor. Coupling between the continuous (fluid) and discrete (solid) phases is incorporated in the interphase momentum exchange term $\mathbf{I}_{f}$, which is derived from the drag force in Eq. (\ref{NEWTS}):\\
\begin{equation} \label{EXCHANGE}
 \mathbf{I}^k_{f} = \frac{1}{\nu_k} \sum_{i=1}^{N_s^k} \mathbf{F}_d^{i \in k} K(\mathbf{X}_{i\in k},\mathbf{x}_k)
\end{equation}
\noindent This states that the interphase momentum exchange in the $k^{th}$ computational cell is dictated by the particle drag force $\mathbf{F}_d^{i \in k}$ associated with the $i^{th}$ particle residing within the cell with a volume $\nu_k$. The particle force within the cell is modulated by the kernel $K(\mathbf{X}_{i \in k}^i,\mathbf{x_k})$, which weights the force of a particle in the $k^{th}$ cell, located by $\mathbf{X}_{i\in k}$, to the grid node located by $\mathbf{x}_k$. This modulated drag force is then summed over all particles residing in the computational cell, $N_s^k$. Particle drag becomes a sum of contributions from the fluid pressure gradient in the $k^{th}$ cell and drag due to the relative velocity between the solid and fluid phases
\begin{equation} \label{FD}
    \mathbf{F}_d^{i \in k} = - \mathbf{\nabla} P_f(\mathbf{x}_k) \nu_s + \beta_s^k\frac{\nu_s}{\epsilon_{s}} \left(\mathbf{V}_f \left( \mathbf{X}_{i\in k}\right)-\mathbf{V}_i \right)
\end{equation}
\noindent where $\mathbf{\nabla} P_f(\mathbf{x}_k)$ is the cell centered fluid pressure gradient, $\nu_s$ is the volume of the solid phase particle, $\epsilon_{s}$ is the solid phase volume fraction, $\beta_s^k$ is the momentum exchange coefficient for all solids residing in the $k^{th}$ cell, and $\mathbf{V}_f(\mathbf{X}_{i\in k})$ is the interpolated mean fluid velocity at the particle location. Combining Eqs. (\ref{EXCHANGE}) and (\ref{FD}), the momentum exchange between the fluid and solid phase becomes:
\begin{equation} \label{EXHCHANGE_PT2}
    \mathbf{I}_{f}^k = -\epsilon_{s} \mathbf{\nabla} P_f(\mathbf{x}_k) + \frac{1}{\nu_k} \beta_s^{k} \frac{\nu_s}{\epsilon_{s}} \left(\mathbf{V}_f \left( \mathbf{X}_{i\in k}\right)-\mathbf{V}_i\right) K(\mathbf{X}_{i\in k},\mathbf{x}_k)
\end{equation}
$\beta_s^k$ is calculated using the model outlined in \citeA{Gidaspow}. The expression for the exchange term is dependent on the fluid volume fraction in the cell
\begin{eqnarray}
  \label{GID}
  \beta_s^k &=&  \cases{150\frac{(1-\epsilon_{f})^2 \eta_f}{\epsilon_f d^2} + 1.75 \frac{\rho_f |\mathbf{V}_f(\mathbf{X}_{i\in k})-\mathbf{V}_i|\epsilon_f}d,&if \( \epsilon_f < 0.8 \)\cr 0.75 C_D \frac{\epsilon_f |\mathbf{V}_f(\mathbf{X}_{i\in k}) - \mathbf{V}_i|}{d} \epsilon_f^{-2.65},&if \( \epsilon_f>0.8 \)\cr} \nonumber \\
\end{eqnarray}
\noindent where $d$ is the particle diameter and $\eta_f$ is the fluid phase viscosity. Thus, the momentum exchange scale factor is a function of solid and fluid phase properties, the relative average velocities between the solid and fluid phases, and the drag coefficient $C_D$, which itself is a function of the cell's particle Reynolds number $Re^k$, a non-dimensional number relating the ratio of fluid-inertial to viscous forces:
\begin{eqnarray} \label{CD}
    C_D &=& \cases{\frac{24}{Re^k} \left(1+0.15 (Re^k)^{0.687} \right),&if \( Re^k<1000 \)\cr 0.44,&if \( Re^k>1000 \)\cr} 
\end{eqnarray} 
This correlation between the drag coefficient and the Reynolds number works to reproduce the standard drag curve relating a sphere's drag coefficient to the Reynolds number (e.g., \citeA{Bird1960}). Note that our simulations are within the dense regime and never reach conditions such that $\epsilon_f>0.8$.  

To relate the resolved dynamics of individual particles to macroscopic flow properties we employ coarse-graining techniques (CG). CG is a computational tool to derive continuous fields from discrete data. We utilize a CG technique developed by \citeA{Fullard2019} and \citeA{Breard2020} following the effort of \citeA{Weinhart2013} (see methods and references, therein, and Supporting Information Section S1 for more information on parameters relevant for scaling).

\subsection{Scaling} \label{SCALING}
This study is concerned with the unit problem of relating basal forces to macroscopic characteristics of granular flows. Due to computational restraints, the simulation setup (described in Section \ref{EXsetup}) is orders of magnitude smaller than real-scaled geophysical mass flows. Nevertheless, analyzing these flows in terms of non-dimensional parameters can provide first-order constraints on the physics that govern mass flow. Though the absolute values of parameters used in our simulations differ from those of real-world particles and flows, non-dimensional parameters allow us to examine the ratios of stress states and timescales such that rudimentary conclusions on momentum transport, rheology, and the overall self-similarity in character of a range of granular flows can be attained.

First, we define the effective friction coefficient, or the ratio of shear to normal stresses, as
\begin{equation} \label{FRICT}
\mu = \frac{|\sigma_{xy}|}{P_s}
\end{equation}
where $|\sigma_{xy}|$ is the domain integrated (coarse-grained) shear stress, and the $P_s$ is the solids pressure (or the pressure generated by the collection of particles that acts upon itself). This friction coefficient is similar to, though is an emergent parameter of, the particle friction coefficient that is assigned to each particle ($\mu_p$). The effective friction coefficient $\mu$ is a constitutive property defined for an entire collection of particles that is a constant of proportionality relating shear stresses to the normal stresses \cite{Midi2004,daCruz2005}. In dense granular flows, $\mu$ is itself a function of the non-dimensional parameter referred to as the inertial number:
\begin{equation} \label{I}
    I = \frac{\dot{\gamma}d}{\sqrt{P_s/\rho_s}}
\end{equation}
where $\dot{\gamma}$ is the coarse-grained shear rate. The inertial number is a quantity that defines the `inertial state' of a granular flow, which in turn has been used as a proxy for flow state or regime. This non-dimensional parameter can be seen as a ratio of the microscopic timescale for particle rearrangement under a confining pressure ($\tau_{micro}=d/\sqrt{P_s/ \rho_s}$), and the timescale for macroscopic deformation as a granular bed is sheared ($\tau_{macro}=1/\dot{\gamma}$) \cite{Midi2004, daCruz2005, Forterre2008}. In this way, at low shear rates and high confining or solids pressures, the inertial number will be low, meaning the timescale for bulk bed deformation is large, and particle motion is relaxed via rearrangement events. This mean that the granular `flow' is characterized by intermittent stick-slip events, rather than bulk failure of the granular bed. 

\citeA{Jop2006} uses the phenomenological relationship between the effective friction coefficient and inertial number to develop a generalized constitutive model for dense granular flows. This model ultimately relates the friction coefficient law $\mu(I)$ to the stress and shear-rate tensors through the use of an effective viscosity defined by $\mu(I)$. In the limit of vanishing shear rates, the model predicts that flow will only occur when the magnitude of shear stresses overcome a fictional yield criterion (e.g., $|\sigma_{xy}| >\mu P_s$). Below this, the collection of grains will behave as a rigid system. This $\mu(I)$ rheology has seen success in modeling dry granular systems and has been been extended to include dense granular suspensions (i.e., multiphase granular flows) \cite{Boyer2011}. This extension utilizes a dimensionless parameter known as the viscous number, expressing the ratio between inertial particle stresses and viscous stresses
\begin{equation} \label{viscous_inert}
    I_v = \frac{\eta_f \dot{\gamma}}{P_s}
\end{equation}
Where $\eta_f$ is the viscosity of the interstitial fluid. This viscous inertial number can be written as a function of the inertial number and the Stokes number $St$, $I_v = I^2/St$, where
\begin{equation} \label{St}
    St = \frac{\dot{\gamma} d^2 \rho_s}{\eta_f}
\end{equation}
The Stokes number is a ratio between the particle timescale characterizing the decay in the particle's velocity due to drag, and the fluid timescale that describes bulk fluid deformation; when $St<1$ particles may act as passive tracers in a background flow field. 
A scaling law unifying $I$ and $I_v$ was introduced by \citeA{Trulsson2012} and written as the modified inertial number $I_m$ by \citeA{Amarsid2017}: 
\begin{equation} \label{modI}
    I_m = \left(\alpha I_v + I^2\right)^{1/2} = I \left(\frac{\alpha}{St}+ 1\right)^{1/2}
\end{equation}
where $\alpha$ is a constant fitting parameter typically of order 1 \citeA{Amarsid2017}. When $St\rightarrow 0$, time scales are dictated by viscous forcing, thus $I_m$ is dominated by the viscous number $I_v$. Conversely, when particles gain inertia, and $St\rightarrow \infty$, $I_m$ reduces to $I$, signifying that dry granular processes control the system. In this study, we have opted to use the dry inertial number $I$, rather than the modified inertial number $I_m$, as our results indicate that, not only does the bulk of our simulations fall into the category where the Stokes number $St$ is greater than one, but that it appears that dry inertial number is sufficient in describing the quantities of interest (Figs. S1 and S2).

Alternative non-dimensional parameters are discussed in the context of debris flows by \citeA{Iverson1997} and can be related to the inertial number $I$ and viscous number $I_v$ \cite{Delannay2017}. The Savage number $N_{Sav}$ is a dimensionless ratio expressing particle inertial stresses and stresses generated by the weight of the flow. If it is assumed that hydrostatic forces are driven by the the solids pressure $P_s$, $N_{Sav}$ is simply the square of $I$:
\begin{equation}
    N_{Sav} = \frac{\rho_s \dot{\gamma}^2 d^2}{(\rho_s-\rho_f) g H} = \sqrt{\frac{\dot{\gamma}^2 d^2 \rho_s}{P_s}} = I^2
\end{equation}
Where $\rho_f$ is the density of the fluid phase (if present), $g$ is the gravitational constant, and $H$ is the thickness of the granular flow. Large-scale sand gravel debris-flow flume experiments conducted by the USGS have estimated values of $N_{sav}\sim0.1$ \cite{Iverson1997}. Similarly, the viscous number $I_v$ and inertial number $I$ can be related to the Bagnold number $N_{Bag}$
 \begin{equation}
     N_{Bag} = \frac{\Phi \rho_s \dot{\gamma} d^2}{(1-\Phi)\eta_f\dot{\gamma}}\sim \frac{I^2}{I_v} = \frac{\dot{\gamma}d^2\rho_s}{\eta_f}
 \end{equation}
 Where $\Phi$ is the concentration, or solids volume fraction, of particles. The Bagnold number $N_{Bag}$, is similar to the viscous number $I_v$ in that is expresses the ratio of particle inertial stress to viscous fluid stresses. $N_{Bag}$ estimates for large-scale debris-flow flume experiments are on the order of $\sim 10^2$ \cite{Iverson1997}. Table 1 expresses the range of non-dimensional parameters for our simulations.

\subsection{Simulation Setup} \label{EXsetup}
We perform two classes of simulations in disparate configurations: monodisperse plane-shear and monodisperse inclined-slope flows. Table \ref{TAB1} provides a summary of simulation parameters.

\begin{figure}[h!]%
\centering
\includegraphics[width=0.6\textwidth]{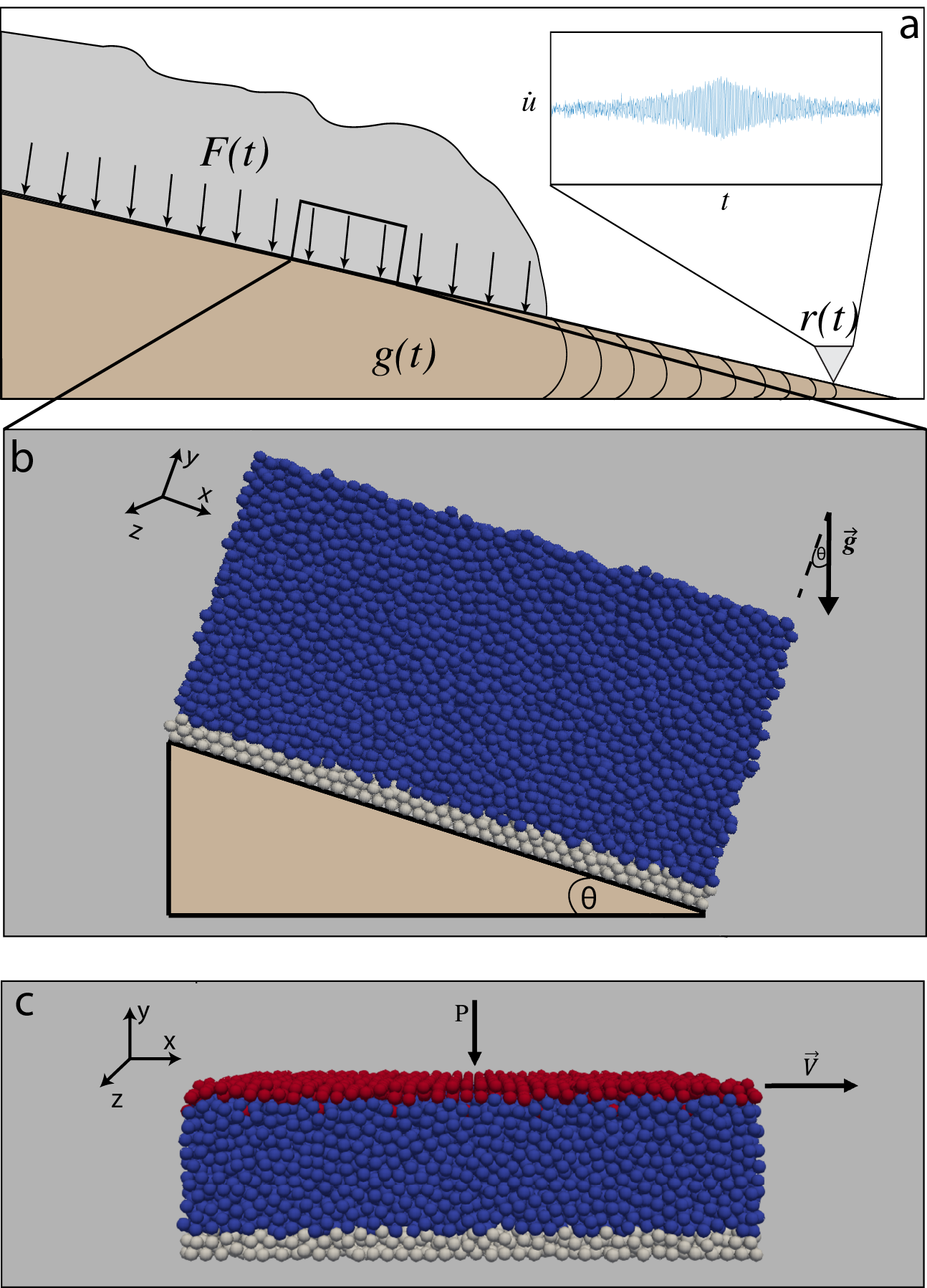}
\caption{(a) Cartoon depicting radiation of elastic energy to receiver via forces exerted by the flow on the substrate and (b-c) renderings of simulation configurations addressing the unit problem of bed forces exerted by granular flows: (b) inclined-slope flow where gravity is rotated by an angle $\theta$ to simulate gravitational forces on a slope; (c) plane-shear configuration, where red particles act as a coherent plate, imposing a constant confining pressure and translational velocity on the blue particles. Gray particles on the bottom of the Y-Z plane act as a static rough and frictional `force plate'.}\label{SETUP}
\end{figure}

In plane-shear configurations, a bed of particles is confined between two plates made frictional and bumpy, being composed of particles that comprise the bed (Fig. \ref{SETUP}(b-c)): the lower plate is made static and acts as a force plate (with an RMS surface roughness $S_{a}\approx0.6d$; Supporting Information S3) recording particle forces (see Section \ref{WALL}); the top plate imposes a constant confining pressure and translational velocity, shearing the bed of grains and initiating flow. The top plate is rigid and contains particles detached downwards from the main body of the plate (Fig. \ref{SETUP}c). This keeps the flow particles from reaching crystallized states. We keep a constant confining pressure of $2000\ Pa$, allowing the height of the bed to evolve due to Reynolds dilation. The relatively low confining pressure, compared to meter scale geophysical flows, will have an effect on the inertial number reached at a given shear rate (see denominator of Eq. \ref{I}). Nevertheless the physical processes at a given flow state described by the inertial number will remain the same. Periodic boundary conditions are imposed in the stream (x) and span-wise (z) directions, with both directions having dimensions equal to $40d$, this is chosen to ensure that simulation dimensions are larger than contact networks that occur in low shearing conditions \cite{Sun2010}. In this configuration, we conduct two categories of simulations, pure granular flows (particles sheared in a vacuum) and flows submerged in two types of fluids: one with a density of $1000\ kg/m^{3}$ and viscosity $8.90 \times 10^{-4}\ Pa\ s$, and a more viscous liquid with the same density and a viscosity of $0.01\ Pa\ s$. The more viscous liquid was chosen to be an order of magnitude greater than water due  to the viscosity of sediment-laden water in debris flow mixtures being 10 to 100 times greater than water \cite{Iverson1997}. Within these two categories (i.e. fluid submerged and dry granular flow), we run two sets of simulations, changing the bed composition from a monodipserse composition of $500\ \mu m$ to a composition of $5\ mm$ particles, while keeping the density constant at $1050\ kg/m^{3}$. This density value provides a slight negative buoyancy in the presence of the interstitial fluid, and is motivated by the density of volcanic tephra \cite{Shipley1983}. We vary the top plate velocity such that the simulations span global shear rates from $10^{-2} - 10^{3}\ Hz$, developing creeping to collisional granular flows. Particle friction is held constant across the simulations at $\mu_p = 0.53$. See Supporting Information Movie S1 for an example of plane-shear simulations in the quasi-static, intermediate, and collisional regimes described by the inertial number ($I\sim10^{-3}$, $I\sim10^{-2}$, and $I\sim10^{-1}$, respectively). 

Inclined plane simulated-flow configurations are conducted similarly, being composed of monodipserse beds placed on top of a frictional and bumpy (force) plate. The bed of particles is first allowed to settle under gravity on a horizontal plane. Then, the gravity vector is rotated to an angle $\theta$ such that $\theta>\theta_s$, where $\theta_s$ is the angle of repose, and the configuration is destabilized. After bed failure, the gravity vector is either rotated to greater values, or decreased below the angle of repose dictated by particle friction, utilizing the hysteretic nature of granular materials \cite{Pouliquen1999}. In this way, inclined-slope simulations reach shear rates from $\sim 10^{-1} - 10^{2}\ Hz$. In these configurations, we maintain a minimum bed height of $20d$, avoiding scale effects for thin flows \cite{Pouliquen1999}. 

\begin{table}[h]
\caption{Summary of simulation parameters.}\label{TAB1}
\centering
\begin{tabular} {lcc}
\textit{parameter} & \textit{variable} & \textit{value} \\
\hline
\textbf{Solid Parameters} \\
\\
diameter$^1$ [m] & $d$ & $0.005,\ 0.0005$\\ 
density [$kg\ m^{-3}$] & $\rho_s$ & $1050$ \\ 
elastic coeff. [$Pa\ m$] & $k_n$ & $(2\times10^{8})d$ \\
angle of repose [$^{\circ}$] & $\theta_s$ & $\approx 20$\\
particle friction coeff. & $\mu_p$ & $0.53$\\
restitution coeff. & $e_n$ & 0.6\\
\hline
\\
\textbf{Fluid Parameters} \\
\\
density [$kg\ m^{-3}$] & $\rho_f$ & $1000$\\
viscosity$^{1,2}$ [$Pa\ s$] & $\eta_f$ & $8.9\times10^{-4},\ 10^{-2}$\\
\hline
\\
\textbf{Simulation Parameters} \\
\\
domain size [$m$]  & $X \times Y \times Z$ & $40d \times \geq 20d \times 40d$ \\
confining pressure$^3$ [$Pa$]  & $P_p$ & $2000$ \\
top plate velocities$^{1,3}$ [$ms^{-1}$] & $V_p$ & $0.0001-16$ \\
inclination angle$^{1,4}$ [$^{\circ}$] & $\theta$ & $20.5-28$ \\
\hline
\\
\textbf{Numerical Parameters}\\
\\
DEM solver timestep [$s$] & $dt_s$ & $\frac{\pi}{50} \left(\frac{k_n}{m_s} - \frac{\eta_n^2}{4m_s^2}\right)^{-1/2}$\\
fluid solver timestep$^2$ [$s$] & $dt_f$ & $(2\times 10^{-2})d$\\
 \hline
 \\
\textbf{Non-dimensional Parameters} & & \textit{simulation range}\\
\\
Inertial Number & $I$ & $10^{-4}-10^{0}$\\
Viscous Number$^2$ & $I_v$ & $10^{-9}-10^{-2}$\\
Stokes Number$^2$ & $St$ & $10^{-3}-10^{3}$\\
Savage Number & $N_{Sav}$ & $10^{-9}-10^{1}$\\
Bagnold Number$^2$ & $N_{Bag}$ & $10^{-3}-10^{4}$\\
 \hline
 \multicolumn{2}{l}{$^{1}$Parameter takes on one of the values listed for a single simulation}\\
  \multicolumn{2}{l}{$^{2}$Parameter relevant when interstitial fluid is present}\\
  \multicolumn{2}{l}{$^{3}$Parameter for plane-shear configuration}\\
    \multicolumn{2}{l}{$^{4}$Parameter for inclined-slope flows}\\
\end{tabular}
\end{table}
 \FloatBarrier
\subsection{Wall Analyses} \label{WALL} 
To link macroscopic flow properties to forcings exerted on the bottom boundary, we record the forces exerted by all particles that are in contact with the bottom rough static plate. In this way, the bottom plate acts as a force plate. We shear the bed of grains until the system reaches a steady state (Fig. S4), and at this point we record forces at this bottom boundary with a frequency of 100 kHz across all simulations. Particle forces are summed across the entire plate, acknowledging that the frequency of this forcing is a function of the size of the numerical force plate \cite{Iverson1997, Jalali2006}: decreasing the area in which particle forces are summed increases the frequency of forcing by the factor at which the plate dimensions are reduced (see Section \ref{REASCALE} for observations and discussion on this effect). From these plate measurements, we calculate the power spectral density (PSD)\\
\begin{equation} \label{PSD}
    P(f) = \frac{N}{\Delta f} |\tilde{F}(f)|^2
\end{equation}
 \noindent where $\tilde{F}(f)$ is the Fourier transform of the total forcing time-series on the lower plate boundary, N is the number of discrete observations (length of the time-series), and $\Delta f$ is the sampling frequency. \citeA{Arran2021} uses the PSD as a measure of the squared fluctuating force. Similarly, we express the fluctuating force as an integrated PSD and normalize it by the time averaged total force (total sum of individual forces) exerted on the bottom plate, squared\\
\begin{equation} \label{Lambda}
    \Lambda = \frac{\int^{f_{ny}}P(f)df}{\left<F(t)\right>^2}
\end{equation}
\noindent where the upper bound of integration is the Nyquist frequency and angled brackets denote time averaging. Effectively, Eq. (\ref{Lambda}) allows us to examine how the bandwidth of forcing spectra changes with flow state, as well as the amplitude of forcing fluctuations relative to the steady-state time-averaged forcing measured by our force plate. Normalization helps encapsulating any magnitude effects that correspond to the bed's weight as well as the momentum transferred from impacts that are fundamentally a function of particle size \cite{Tsuji1992, Yohannes2012}. The amplitude spectrum of the force-plate time series is also used to calculate the mean frequency of the forcing \cite{Vinningland2007, Farin2015}\\
 \begin{equation} \label{FMEAN}
     \bar{f} = \sum \frac{\tilde{A}[f_i]f_i}{\sum{\tilde{A}[f_i]}}
 \end{equation}
 where $\tilde{A}$ is the discrete amplitude spectrum. 
 Forces and velocities at the force plate are used to calculate parameters similar to those described in Section \ref{SCALING}. The effective wall friction coefficient $\mu_w$ is the ratio of the time averaged sum of basal forces in the stream-wise and plate-normal directions \cite{Artoni2015}\\
 \begin{equation} \label{MU_W}
     \mu_w = \frac{\left<F_x(t)\right>}{\left<F_y(t)\right>}
 \end{equation}
Similarly, we compare the CG'd granular (the granular temperature averaged over the entire flow) temperature to the granular temperature at the bottom force-plate wall, defined as the time and spatially averaged mean squared velocity fluctuation of particles in contact with the bottom plate (thus, slip velocities)\\
\begin{equation} \label{WALLGT}
    T_i^{w} = \left<\left(v_i^w-V_i^w\right)^2\right> 
\end{equation}
Where $T_i^w$ is the wall granular temperature in the $i^{th}$ direction, $v_i^w$ is the wall-averaged instantaneous velocity in the $i^{th}$ direction, and $V_i^w$ is the time and wall averaged slip velocity in the $i^{th}$ direction. Note, as with all parameters defined at the force plate, uncertainty increases with shear rate, as the total number of particles in contact with the bottom plate at write-out decreases.
\FloatBarrier
\section{Results}\label{Results}

\subsection{Granular Rheology}\label{Rheo}
\FloatBarrier
\begin{figure}[h!]%
\centering
\includegraphics[width=0.6\textwidth]{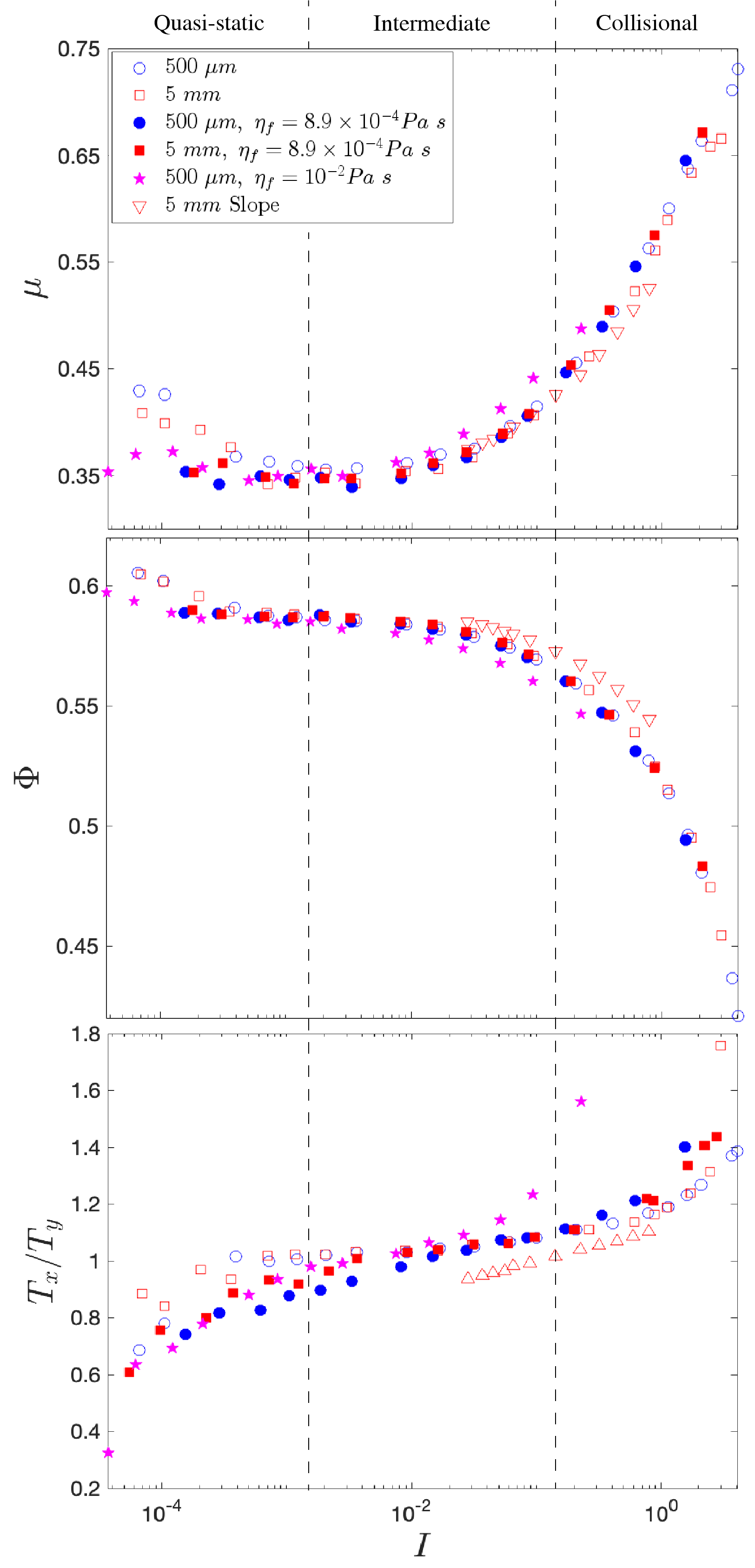}
\caption{Granular rheology described by (a) the effective friction coefficient $\mu$ and (b) the solids concentration $\Phi$, as a function of inertial number $I$, derived from coarse-graining discrete data. Vertical dashed lines outline the three regimes detailed in the text.}\label{RHEO}
\end{figure}
The rheology of granular flows and concentrated suspensions is well established with empirical constitutive laws taking the form $\mu(I)$ and $\Phi(I)$, stating that the effective friction coefficient $\mu$ and the solids volume concentration $\Phi$ can be predicted by the inertial number $I$, with similar laws being expressed with the modified inertial number $I_m$ \cite{Jop2006, Amarsid2017}. Utilizing CG, we have derived bed averaged fields for the stress tensor and concentration, recovering these scaling relationships for monodisperse flows with and without an interstitial fluid phase (Fig. \ref{RHEO}). We have opted to plot data using the dry inertial number $I$, rather than using $I_m$, for reasons outlined in Supporting Information Section S2. 

In Fig. \ref{RHEO}a, we observe regimes in $\mu(I)$ space that have been observed by previous efforts \cite{Midi2004,daCruz2005,Forterre2008,DeGiuli2017,Breard2020, Breard2024}. First, we identify a velocity-weakening regime when $I<10^{-3}$. Though inertial numbers in this range are most commonly referred to as the quasi-static regime (e.g., \citeA{Midi2004, Breard2020}), where $\mu$ stays constant with increases in shear rate, we observe behavior that reflects non-monotonic scaling between $\mu$ and $I$. This has been posited to arise from endogenous noise generated in low shear conditions \cite{DeGiuli2017}. Here, velocity weakening is most prevalent in simulations conducted in the absence of a fluid phase. As non-monotonic $\mu(I)$ scaling may be a special case of granular flows that are characterized by intermittence (i.e., stick-slip flowing behavior; e.g., \citeA{Baldassarri2019}), we have opted to express this regime as concomitant with the quasi-static regime. Moreover, as $I\rightarrow 10^{-3}$, $\mu$ reaches a relatively constant value, which is characteristic of the quasi-static regime. Similarly, at this point, the particle concentration maintains a constant value (Fig. \ref{RHEO}b). Effective friction begins to exhibit rate dependence at inertial numbers between $10^{-3}$ and $10^{-2}$, signifying the transition into what is referred to as the intermediate regime. Here, the flowing bed of grains behave phenomenologically like a liquid; the granular bed is no longer able to support itself under shear loads, resulting in continuous deformation under shear. In the intermediate regime, particles remain in prolonged contact with one another, forming an evolving contact (force-chain) network that rotates, breaks, and reforms itself in time \cite{Majmudar2005, Forterre2008}. In this regime, the particle concentration begins to display dilatancy, or decreases in particle concentration with increasing shear rate (Fig. \ref{RHEO}b). Finally, in the limit when $I>10^{-1}$, the flow enters into the collisional regime (also known as the inertial regime). In this limit, the bed dilates rapidly, prolonged contacts are reduced, and binary particle collisons become the primary mechanism for momentum transfer \cite{Midi2004, daCruz2005, Forterre2008}.

\begin{figure}[h!]%
\centering
\includegraphics[scale=0.25]{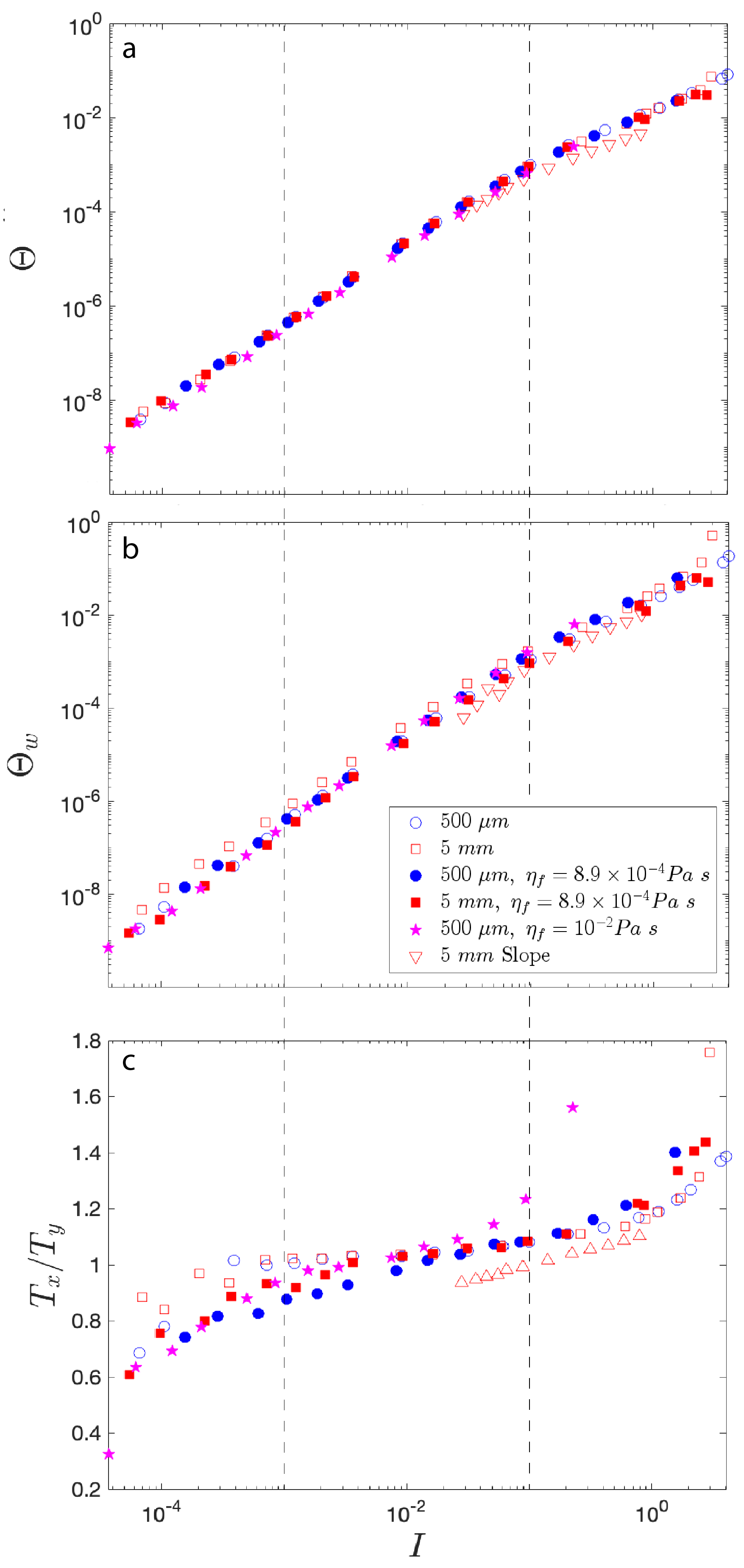}
\caption{(a) Flow-averaged --- coarse-grained --- non-dimensional granular temperature $\Theta$, (b) wall-averaged non-dimensional granular temperature at the force plate, and (c) the ratio of stream-wise and span-wise flow-averaged granular temperature as a function of inertial number $I$. Vertical dashed lines show the three inertial regimes: quasi-static $I$ $<$ $10^{-3}$, intermediate $10^{-3}<I<10^{-1}$, and collisional $I>10^{-1}$.}\label{THETA}
\end{figure}

In conjunction with the ratio of stresses ($\mu$) and concentration ($\Phi$), we can help classify the internal structure of our flows using granular temperature. Fig. \ref{THETA} depicts non-dimensional isotropic granular temperature $\Theta$ ($\Theta = \rho_s T P_s^{-1}$) as a function of $I$, coarse-grained as a bed-averaged quantity ($\Theta$) and wall-averaged ($\Theta_w$) --- taking the average granular temperature of all particles in contact with the force plate (Figs. \ref{THETA}a and \ref{THETA}b, respectively). Both measures of granular temperature display near-linear log-log scaling with $I$, with data points collapsing to a single curve, showing that this relationship is independent of particle size, the presence of a liquid phase, and changes in flow configuration. This suggests that the fluctuating kinetic energy of particles at the base of the flow, those transferring flow energy to the force plate, can be described by the total bed averaged inertial state $I$ (Fig. \ref{THETA}b). 

Granular temperature is commonly assumed to be isotropic in many approaches that utilize kinetic theory applied to rapid granular flows (e.g., \citeA{Johnson1987,Syamlal1993,Berzi2020}. Nevertheless, the anisotropic nature of granular temperature has been acknowledged rather early in the history of granular kinetic theory (e.g., \citeA{Campbell1989}). To show this, we plot the ratio of the stream-wise granular temperature to the component normal to the bottom plate (Fig. \ref{THETA}c). The degree of granular temperature anisotropy gives insight into how momentum is communicated and advected through the system, with isotropic quantities corresponding to fluctuations that arise due to particle contacts, and anisotropy in the stream-wise direction resulting from particle motion normal to shear, likely driven by particle-particle collisions and deflections \cite{Campbell2006}. In general terms, when $I<10^{-3}$, granular temperature is anisotropic in the y-direction (normal to the bottom plate); as $I\rightarrow 10^{-1}$, the stream-wise component begins to exceed the vertical component (Fig. \ref{THETA}c). Note that the viscosity of the interstitial fluid phase plays a large role in this behavior; for example, simulations conducted in a viscous fluid show that the stream-wise component increases to over 20\% of the y-normal component when $I\approx0.1$, where the water submerged and dry flows experience a similar increase when $I\approx0.8$ and $I\approx1$, respectively. The directionality of granular temperature seems to be related to the regimes described by $\mu$ and $\Phi$ in Fig. \ref{RHEO}: anisotropy in the y-direction correlates with quasi-static regime when $I<10^{-3}$ and $\mu$ decreases with increasing shear rates, isotropy occurs primarily in the quasi-static and intermediate regimes, and granular temperature becomes anisotropic in the stream-wise direction as flows dilate and approach the collisional inertial regime. 

\begin{figure}[h!]%
\centering
\includegraphics[width=0.6\textwidth]{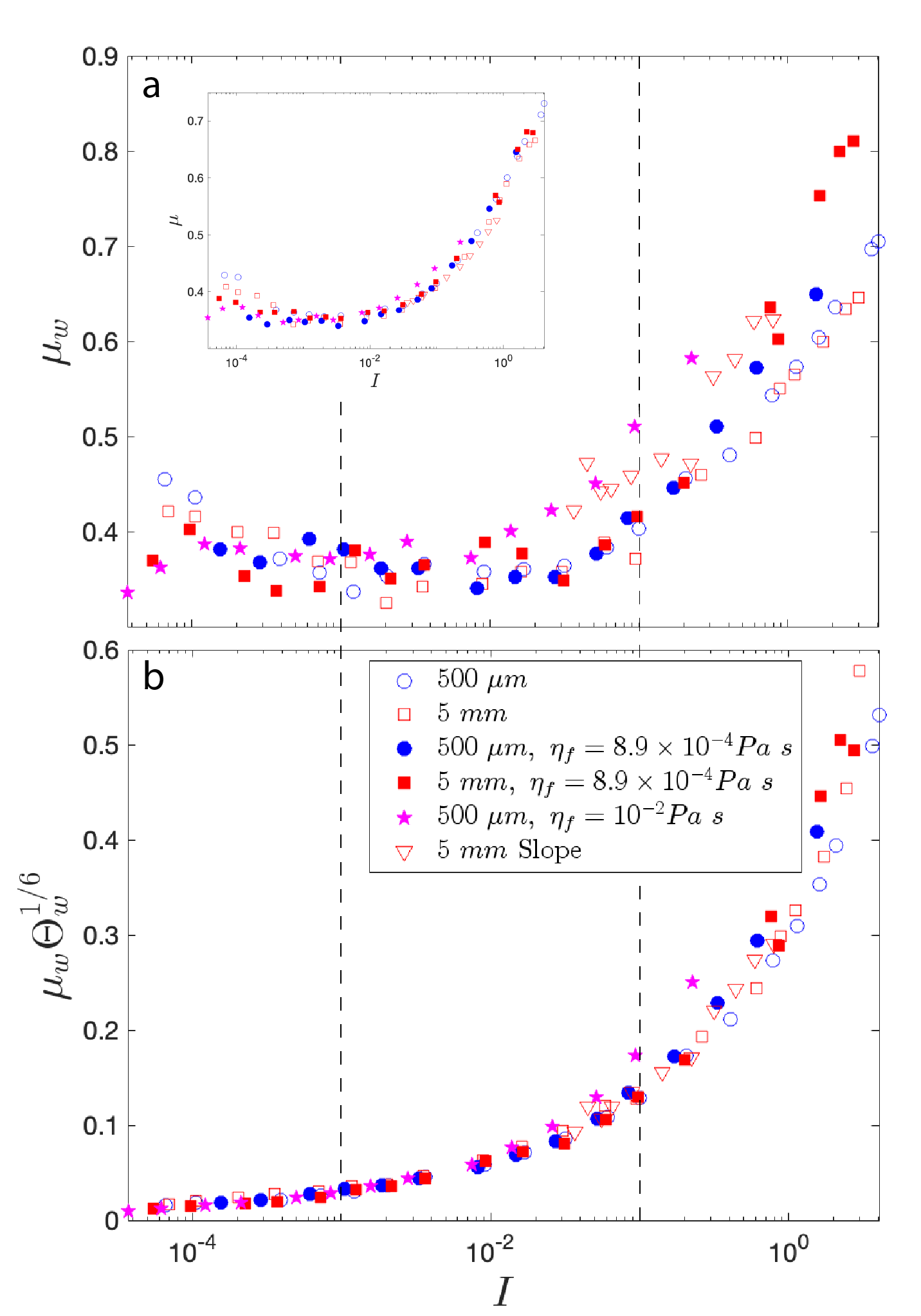}
\caption{(a) Wall-averaged (bottom boundary flow-plate interface) effective friction coefficient $\mu_w$, with inset showing the bed-averaged friction coefficient and (b) wall-averged friction coefficient scaled by the wall-averaged non-dimensional granular temperature $\Theta_w$, proposed by \citeA{Kim2020}, as a function of inertial number $I$. Vertical dashed lines demarcate the quasi-static, intermediate, and collisional inertial regimes}\label{WALL_mu}
\end{figure}

Figs. \ref{RHEO} and \ref{THETA} show that granular flow rheology and internal energy fluctuations, averaged over the entire flow and at the force plate, scale with a macroscopic description of flow inertial state. We can continue to explore the relationship between averaged flow states and basal interactions by examining the effective friction coefficient at the force plate, or effective wall friction, $\mu_w$ (Fig. \ref{WALL_mu}). $\mu_w$ scales very similarly to the bed-averaged friction coefficient, but with much more spread: the general trend of friction decreasing at low inertial states, transitioning to a relatively constant value, then increasing with increases in $I$, is maintained. Nevertheless, there is much more spread amongst the simulations, showing that changing the flow configuration, and including the presence of an interstitial fluid, causes deviations away from a universal $mu_w(I)$ scaling. \citeA{Kim2020} suggest that the friction coefficient rescaled by non-dimensional granular temperature $\Theta$ helps collapse variability across flow configurations. Following this, Fig \ref{WALL_mu}b depicts their suggested scaling, but we instead use quantities defined at the force plate. This shows that variability in the ratio of wall stresses, tracked by the flow-averaged inertial number, is effectively balanced by energy fluctuations at the bottom boundary. 
\FloatBarrier

\FloatBarrier
\subsection{Wall forces} \label{WALL_FORCES} 
Wall forces (or basal stresses) are how geophysical mass flows effectively communicate flow momentum to the earth. Encoded in this exchange is an energy cascade that converts flow driving energy to shear work. This work is ultimately dissipated through particle-particle and particle-boundary interactions, as well as frictional heating \cite{Iverson1997}. To this end, we will examine how simulated wall forces --- or basal forces exerted by the flow on to the rough and static bottom plate --- fluctuate, using the power spectral density (PSD) of the basal forcing time series. These spectra will be used to relate the power of these fluctuations back to macroscopic rheologic behavior observed in Section \ref{Rheo}. 

\begin{figure}[h!]%
\centering
\includegraphics[width=\textwidth]{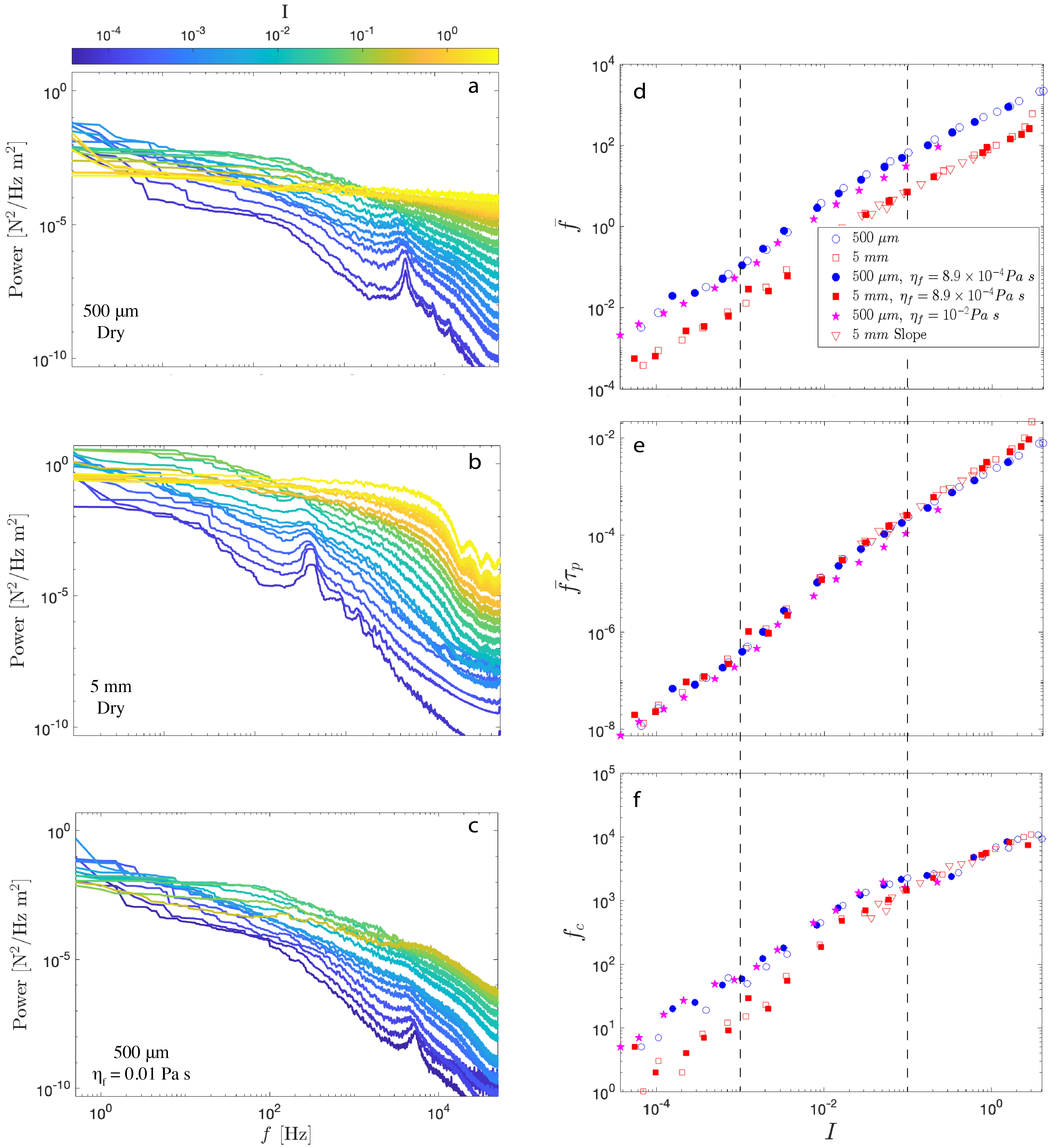}
\caption{(a-c) RMS basal-force power spectral densities (colored by inertial number) for dry simulations with $500\ \mu m$ (a) and $5\ mm$ particles (b), and simulations with an interstitial viscous fluid with $500\ \mu m$ particles (c).The right column depicts (d) mean frequency $\bar{f}$ in $Hz$ of the forcing spectra, (e) $\bar{f}$ rescaled by the particle response timescale $\tau_p$ , and (f) corner frequency $f_c$ as a function inertial number $I$. Veritcal dashed line in (d-f) depict the three inertial regimes: quasi-static, intermediate, and collisional.}\label{F_TIME}
\end{figure}

Fig. \ref{F_TIME}a-c show the root-mean-square PSDs derived from the sum of wall forces measured by the force plate for dry simulations containing $500\ \mu m$ particles, $5\ mm$ particles, and simulations with $500\ \mu m$ particles submerged in a viscous fluid (Figs. \ref{F_TIME}a, \ref{F_TIME}b and \ref{F_TIME}c, respectively), with each spectrum colored by coarse-grained inertial number $I$. As $I$ increases, the PSD for all simulations begins to lift, transitioning from a relatively constant linear decrease in power with frequency at low inertial states, to a near-flat response, up to a corner frequency, when $I\sim1$. PSDs have been scaled to account for changes in the size of the simulated-flow configuration and the subsequent effect on the absolute quantities measured in the PSD. We scale by the plate area, using the common assumption that basal forces are random and uncorrelated, such that the amplitude spectra of these forces will scale with the square root of the number of forcing locations (e.g., \citeA{Tsai2012,Farin2019,Allstadt2020}). Nevertheless, we must note that the assumption of uncorrelated forcing breaks down in inertial states where $I<10^{-1}$, when forces are correlated in space, and definite length scales are intrinsically related to flow rheology \cite{Lois2007,Ge2024}. 

Figs. \ref{F_TIME}d and \ref{F_TIME}e show the mean frequency $\bar{f}$ of each simulation, and the mean frequency scaled by the particle deformation timescale $\tau_p$, a function of particle size, density, and normal elastic coefficient:
\begin{equation} \label{tau_p}
    \tau_p = \left(\frac{d}{\sqrt{k_n/\rho_s d}}\right) 
\end{equation}
Fig. \ref{F_TIME}d shows that the mean frequency is a monotonic increasing function of $I$, ranging from less than $10^{-3}\ Hz$ to slightly greater than $10^{3}\ Hz$, depending on particle size. This reflects the fact that the character of the PSDs change with inertial number: as $I$ increases, more power is relegated to higher frequencies, causing the spectra to lift from the log-log linear decay (Fig. \ref{F_TIME}a-c). The particle size dependence in $\bar{f}$ is completely captured by $\tau_p$ thus, implicitly, particle scale properties such as particle size and and elastic coefficient $k_n$ (Fig. \ref{F_TIME}e). Meaning as particle size decreases by an order of magnitude, the mean frequency measured at a given inertial number will increase by an order of magnitude. This results in the PSDs of the $500\ \mu m$ flows being shifted to the right in power-frequency space relative to the $5\ mm$ flows (Figs. \ref{F_TIME}a and \ref{F_TIME}b). This behavior is similarly exhibited in Fig. \ref{F_TIME}f where the corner frequency $f_c$ (defined as the frequency at which the signal has dropped below 50\% of the mean power) is shown as a function of inertial number. At low inertial numbers ($10^{-4}<I<10^{-2}$) the corner frequency $f_c$ of each configuration seems to be dictated by particle size (i.e., $f_c$ values differ by an order of magnitude when the particle size changes by an order of magnitude). Nevertheless, as $I$ increases, the two curves that were once dictated by particle size meet just as $I>10^{-1}$, meaning that once the flows enter into the collisional regime, the corner frequency is controlled solely by the inertial state of the flow (Fig. \ref{F_TIME}f). Note that the concavity of the curves shown in Figs. \ref{F_TIME}d-f change as the inertial number approaches, and becomes greater than, $10^{-2}$, providing evidence that changes in flow regime, as the flow enters into the intermediate inertial state, is effectively captured in forces recorded by our numerical force plate.

\begin{figure}[h]%
\centering
\includegraphics[width=\textwidth]{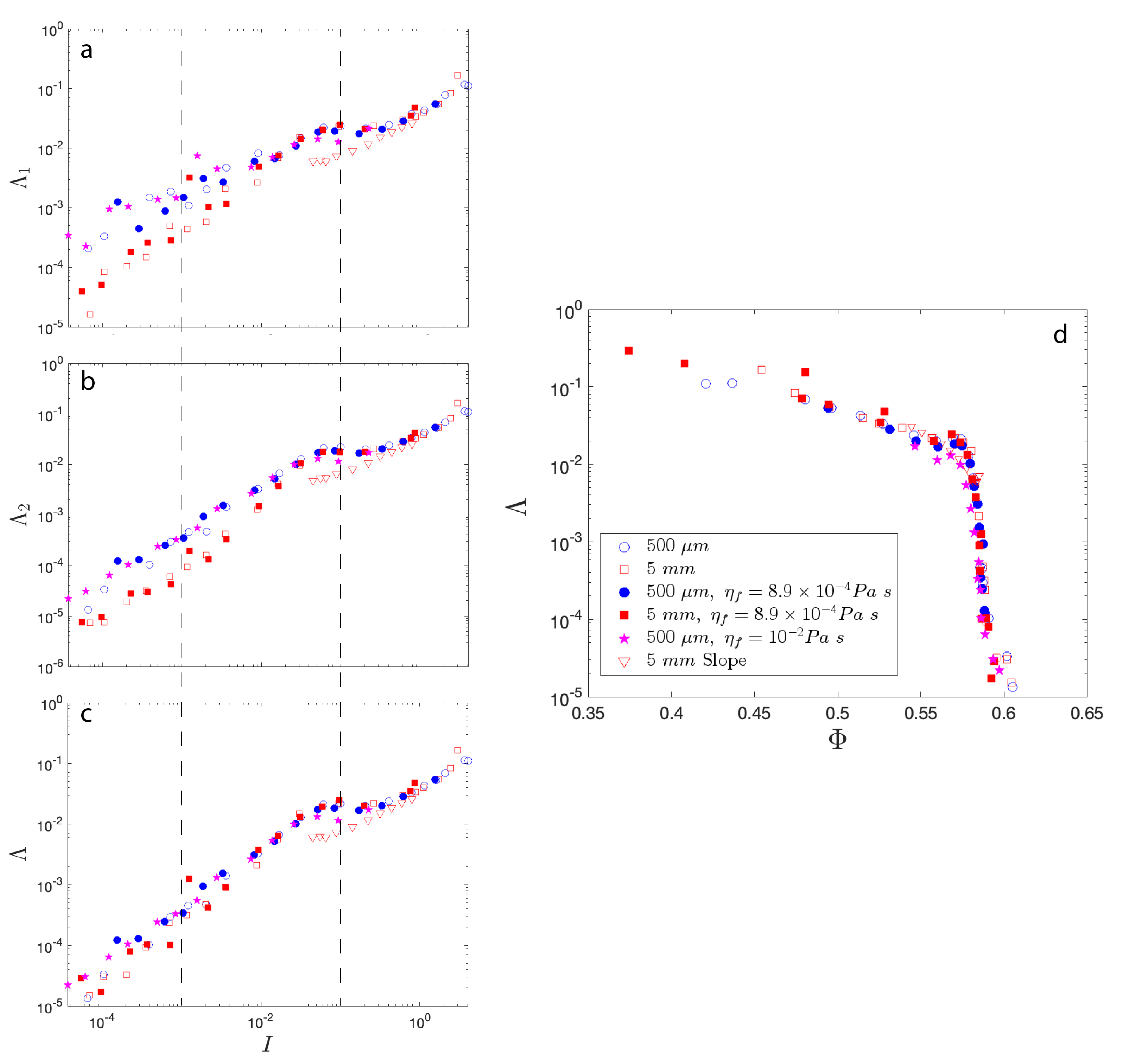}
\caption{(a-c) Non-dimensional forcing fluctuations $\Lambda$ as a function of inertial number $I$, with vertical dashed lines inertial regimes, and (d) as a function of particle concentration $\Phi$. (a-c) show the effect of changing the bounds of integration in Eq. (\ref{Lambda}): $\Lambda_1$ an integral with the lower bound being $1\ Hz$ and the the upper being $f_{ny}$ (a), $\Lambda_2$ with a lower bound being $10\ Hz$ and the upper being $f_{ny}$ (b), and $\Lambda$ showing the integral with variable lower bounds dictated by particle size, $1\ Hz$ for $5\ mm$ particles and $10\ Hz$ for $500\ \mu m$ particles (c). (d) Depicts $\Lambda$ (variable lower bounds, c) as a function solid particle fraction $\Phi$, with critical shift in scaling occurring at $\Phi_c \approx 0.57$}\label{Lambda_I}
\end{figure}

This study is in part motivated by the ability of granular flows to radiate elastic energy that can be observed in real-time in the form of seismic signals (see Allstadt et al., 2018 for a review on signals used to observe geophysical mass flows). A static forcing will not generate an observable signal (i.e., if one merely stands still next to a seismometer, no ground movement will be recorded). Therefore, we are concerned with the basal (wall) force fluctuations in-time. To this end, we plot the non-dimensional integral of the simulated PSDs $\Lambda$ --- as defined in Eq. (\ref{Lambda}) --- allowing a glimpse at how the variance in the bed-force time series (measured by the PSD), and the binning of power in frequency space, changes with changes in inertial state (Fig. \ref{Lambda_I}a-c). Eq. (\ref{Lambda}) is defined as the integral over the PSD, and in Figs. \ref{Lambda_I}(a-c) we show the effect the lower bounds of integration has on $\Lambda$ (always excluding the zero-frequency mean): Fig \ref{Lambda_I}a shows $\Lambda_1$, the integral from $1\ Hz$ to $f_{ny}$; Fig \ref{Lambda_I}b shows $\Lambda_2$, the integral from $10\ Hz$ to $f_{ny}$; and finally, Fig \ref{Lambda_I}c shows $\Lambda$ as function of $I$, with the integral having variable lower bounds of integration that is dictated by the particle size: $10\ Hz$ for simulations using $500\ \mu m$ particles and $1\ Hz$ for simulations using $5\ mm$ particles. Fig. \ref{Lambda_I}d depicts this choice in $\Lambda$ (i.e., lower bounds of integration dictated by particle size), as a function of the solid volume concentration $\Phi$. We have chosen to use this definition of $\Lambda$ as Fig. \ref{F_TIME} shows that, at equivalent inertial states, the relative binning of power in frequency space is shifted by an order of magnitude when the particle size changes by an order of magnitude. Thus, as we are concerned with how basal forces record bulk rheology, rather than effects that are dictated by particle size, we have opted to change the lower bounds of integration based on particle size to reflect the relative decrease in high frequency energy when particle size increases. Here, we are establishing a theoretical framework for what flow information can be effectively encoded in basal forcing signals, future work is needed to examine how this can be generalized to polydipserse flows.

Regardless of the bounds of integration, and when the zero frequency mean is excluded from the integral, the character of the $\Lambda$ vs $I$ curves remain the same: under plane-shear conditions, spread in $\Lambda$ lessens as $I$ approaches and becomes greater than $10^{-2}$, or as flows are well within the intermediate regime described in Section \ref{Rheo}. This scaling persists up until the threshold of the inertial collisional regime, where $I\sim10^{-1}$. Here, $\Lambda$ briefly shifts from an monotonic increasing function of $I$, and decreases with $I$, until finally increasing again as a log-log linear function as $I>10^{-1}$. Previous work has observed a similar correlation between the integral of the PSD and an inertial number estimate \cite{Arran2021}. Here we extend this view to a much larger range of inertial states, analyzing this over a wider frequency band and comparing it against an inertial number that is estimated via our coarse-grained stress tensor. Further, we observe that $\Lambda$ is a strong function of solid concentration $\Phi$, with a shift in scaling behavior when $\Phi = \Phi_c \approx 0.57$ (Fig. \ref{Lambda_I}e). A related correlation has been observed in transient flume experiments utilizing sediment-water mixtures \cite{Piantini2023}. Importantly, Fig. \ref{Lambda_I}e shows a strong functional connection between basal forcing and flow concentration, an important parameter for unifying non-local and local rheologies observed in disparate granular systems \cite{Breard2024}. 


\FloatBarrier
\section{Discussion} 
Geophysical surface flows are complex phenomena that pose challenges to life and property. Near-field and along-channel seismometers provide indirect data related to these events, having potential use as early warning devices and flow monitors. Nevertheless, inverting seismic signals generated by surface mass flows is made difficult by site specificity, and a lack of an observational framework that relates basal forces to bulk flow properties. While phenomena such as debris flows, lahars, and pyroclastic density currents are highly complex and differ greatly in phenomelogical expression, they share an underlying nature of multiphase, in the case of non-negligible fluid-drag, granular mixtures \cite{Iverson1997,Delannay2017,Lube2020}. While the simulations produced in this study are highly idealized monodisperse flows, composed of perfect spheres in simple-shear and inclined-slope configurations that have attained steady state, we believe these simulations capture the first order effects of granular-stresses exerted onto the boundary. In this way, we seek to relate the rich granular rheology described by, e.g., \citeA{Midi2004}, \citeA{daCruz2005}, \citeA{Forterre2008}, and \citeA{Boyer2011} to basal tractions generated by geophysical granular flows. Our data suggests that bulk flow properties, such as inertial state, can effectively be encoded in forces transmitted to the boundaries. Further, a phase space defined by the integral of basal forcing power spectra and inertial number demarcates four regimes that hold a relation to flow regimes describing the states of granular matter described in Section \ref{Rheo}. This relationship may be leveraged to model basal forces via a sub-grid parameterization that can be used in models that simulate flows on real-world scales, but whose resolution precludes them from directly capturing granular flow-boundary stresses. Future efforts can continue to paint a fuller picture of bulk flow rheology and basal forces, diverging from the ideal explored here, and considering broader particle size distributions, dilute particle concentrations, and non-spherical particles.  

Below, we will begin our discussion by considering granular temperature as a measure of internal fluctuating kinetic energy, originating from shear-work and flow driving forces. We will examine how granular temperature changes across inertial space, as well as how it is affected by flow configuration. We will then consider how these changes are related to particle-scale contacts, ultimately leading to the emergent flow-scale evolution of continuous fields describing our flows. This will allow us to link observations of bulk granular rheology described in Section \ref{Rheo} to basal forcing spectra described in Section \ref{WALL_FORCES}.

\subsection{Fluctuating energy generated through shear work, and enhancement of anisotropy through fluid-solid coupling}

Granular temperature is a measure of velocity fluctuations that has been used widely through the development of granular kinetic theory. This temperature is traditionally used to infer pressure, shear stresses, and viscosity in high-energy, dilute systems \cite{Lun1984, Johnson1987}. Granular temperature may play an important role governing the runout of natural granular flows \cite{Campbell1989}. For example, granular temperature is observed to concentrate in dilute regions below dense cores of high speed unidirectional flows, reducing the effective basal friction with increasing mass \cite{Brodu2015}. Granular temperature, when thought of as the fluctuating energy of a particle, is an integral part of the energy cascade inherent to granular flows as dissipative systems. In this cascade, flow driving energy is converted into macroscopic kinetic energy, as described by the mean velocity, which in turn drives velocity fluctuations of individual grains. These velocity fluctuations lead to dissipative interactions, closing the cascade of energy loss to irrecoverable particle deformation and frictional heating \cite{Campbell2006}. Here, we report granular temperature as derived from the coarse-grained kinetic stress tensor, a continuous field originating from the direct measure of discrete particle velocity fluctuations. 

In these simulations, granular temperature is produced through shear work and particle-particle interactions --- rather than boundary layer effects resulting in wake structures contributing to kinetic energy fluctuations \cite{Mehrabadi2015}. Figs. \ref{THETA}a,b show that, when treated as an isotropic quantity, even when fluid drag outpaces particle inertia, i.e., when $St<1$ (Fig. S1), non-dimensional granular temperature $\Theta$ is well parameterized by the dry inertial number $I$. This view changes in light of Fig. \ref{THETA}c, which shows that the degree of anisotropy varies with $I$ and interstitial fluid composition. Anisotropy in granular temperature has in the past been related as a measure of anisotropy in the transmission of momentum, as well as the mode of granular temperature production, i.e., through particle-particle interactions or shear work and advection through velocity gradients \cite{Campbell2006}. From the previously defined intermediate regime up to the collisional inertial limit (or when $10^{-3}$ $<I$ $<10^{-1}$), macroscopically defined granular temperature is roughly isotropic, suggesting that this variable is maintained through particle-particle interactions, leading to an equal partitioning of fluctuating energy in three dimensions. As $I\rightarrow 10^{-1}$, granular temperature becomes anisotropic in the x direction, suggesting that temperature production becomes dominated by velocity fluctuations arising from particles advecting/deflecting normal to the velocity gradient. As shear induced velocity gradients develop, random particle motion normal to the gradient can enhance the generation of granular temperature; this is known as the streaming effect \cite{Campbell2006}. The streaming effect causes granular temperature to be anisotropic in the shearing direction. Since viscous forces contribute to particle-shearing, the inertial number at which this anisotropy occurs is a function of the presence of, and viscosity of, the interstitial liquid. Note that the dry granular simulations are lacking these viscous shear contributions, therefore simulations involving a liquid phase reach anisotropic conditions at lower inertial numbers, compared to the simulations without a liquid phase (Fig. \ref{THETA}c). At the low inertial number limit, when $I<10^{-3}$, granular temperature is anisotropic in the direction normal to the bottom boundary. This occurs when the effective friction coefficient $\mu$ and particle concentration $\Phi$ decrease with $I$ (Fig. \ref{RHEO}). With this in mind, anisotropy at these low inertial numbers can be linked to particle reorganization; particles mount those that reside in layers lower and adjacent to them, sliding and falling back down under the force of the top plate's confining pressure \cite{DeGiuli2017}. Moreover, bed spanning force chains in this low inertial number regime are long lived, but susceptible to buckling events, causing motion normal to the shearing plane during slip events \cite{Tordesillas2011}. Individual contact forces further reflect this directional dependence on inertial number (Fig. S5).

The differences in the two modes of granular temperature production (particle-particle interactions or shear induced `heating') can also be indicated through the relationship between granular temperature and inertial number. Fig. \ref{THETA} shows a direct relationship between inertial number and granular temperature rescaled by the solids pressure and the particle density. Rescaling is required for all of the simulations to collapse to a master curve. The relationship between granular temperature and inertial number is sensitive to the solids pressure and density. Thus, the absolute value of granular temperature is a function of flow configuration: the inclined-slope flow simulations reach granular temperatures nearly an order of magnitude less than that of a shear-cell simulation at an equivalent inertial number (Fig. S6). We relate this decrease to the fact that the inclined-slope simulations do not experience the same magnitude of shear, suggesting that the solids pressure corrects for the lack of shear work provided to generate granular temperature. This is similar to the effect produced by rescaling the coarse-grained shear rate by the timescale of particle rearrangement, thus giving equivalent inertial numbers across flow configurations. Non-dimensional granular temperature can help provide a link across not only various flow geometries \cite{Kim2020} but granular matter under disparate confining pressures. This is important for evaluating bed forces as solids pressures may vary significantly in natural flows interacting with complex topography \cite{Benage2016}.

\FloatBarrier
\subsection{Basal forces record granular phase transitions} \label{PHASE} 

Previous laboratory experiments have constrained a correlation between $I$ and an estimate of the variance in basal forces ($\Lambda$) for inclined plane flows at high inertial numbers (i.e., $I$ $>$ $10^{-1}$; \citeA{Arran2021}). Here, we expand this by performing analysis that probes a wider range of inertial states, with $I$ spanning from $10^{-4} - 10^{0}$. With this expansion, we are able to demarcate distinct flow regimes by correlating changes in $\Lambda$ with macroscopic quantities derived from coarse-graining procedures. Fig. \ref{Lambda_I} shows that there are distinct changes in the nature of how $\Lambda$ scales with inertial number, most notably, as the flow transitions from the intermediate, liquid-like flow regime, to the collisional gas-like flow regime. To the author's knowledge, this is a newly discovered regime that expresses itself in the transmission of flow driving forces into boundary stresses. Therefore, we will describe this regime in detail before considering the relationship between $\Lambda$ and $I$ in the three other previously defined flow regimes (i.e., the quasi-static, intermediate, and collisional regimes).

In the shear-cell simulations, as $I\rightarrow10^{-1}$, $\Lambda$ ceases its near constant increase with $I$, plateauing then dropping slightly just as $I>10^{-1}$ (Fig. \ref{Lambda_I}c). This is, to an extent, mirrored by the inclined-slope simulations: $\Lambda$ stays relatively constant with increasing shear rates when $10^{-2}<I<10^{-1}$, then recovers the scaling behavior observed by the shear-cell simulations. This shift is corroborated in Fig. \ref{Lambda_I}d which depicts a critical value of solid concentration $\Phi_c \approx 0.57$ above which $\Lambda$ decreases rapidly from a local minimum. At concentrations greater than $\Phi_c$, the packing fraction (or concentration of solid particles) begins to inhibit strain \cite{Zhang2017}. This is analogous to a critical volume fraction that defines a glass-like transition in which the kinetic properties of a liquid are diminished as the liquid is undercooled and avoids crystallization \cite{Cohen1959}. Moreover, a similar critical value of $\Phi$ is observed in dense suspensions, below which the rheology of the system is controlled by the Stokes number of the solids phase, and above which is a contact dominated regime that is rate-independent up to a critical shear rate \cite{Ness2015}. 

\begin{figure}[h!]%
\centering
\includegraphics[scale = 0.4]{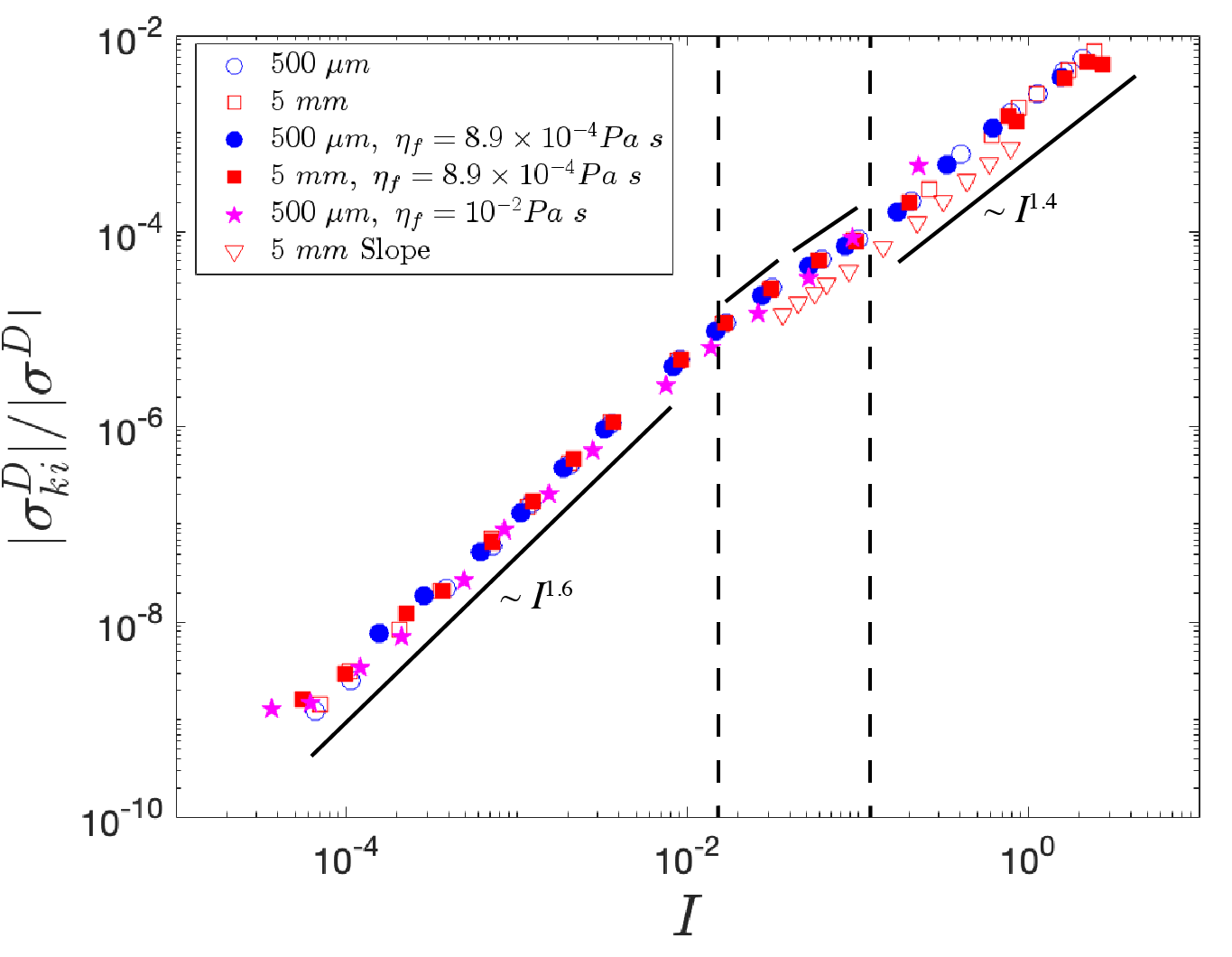}
\caption{ Ratio of the magnitude of deviatoric kinetic stress to the magnitude of total deviatoric stress ($\sigma^{D}$ $=$ $\sigma^{D}_c$ + $\sigma^{D}_{ki}$, where $\sigma^{D}_c$ is the deviatoric contact stress and $\sigma^{D}_k$ is the deviatoric kinetic stresses) as a function of inertial number $I$. Vertical dashed lines demarcate when the concavity of the scaling relationship changes, transitioning from a log-log linear scaling that goes as $I^{1.6}$ (solid line from $10^{-4}<I<10^{-2}$) to $I^{1.4}$ (solid line from $10^{-1}<I<10^{0}$).} \label{SIGMA_SIGMA}
\end{figure}

\begin{figure}[h!]%
\centering
\includegraphics[width = \textwidth]{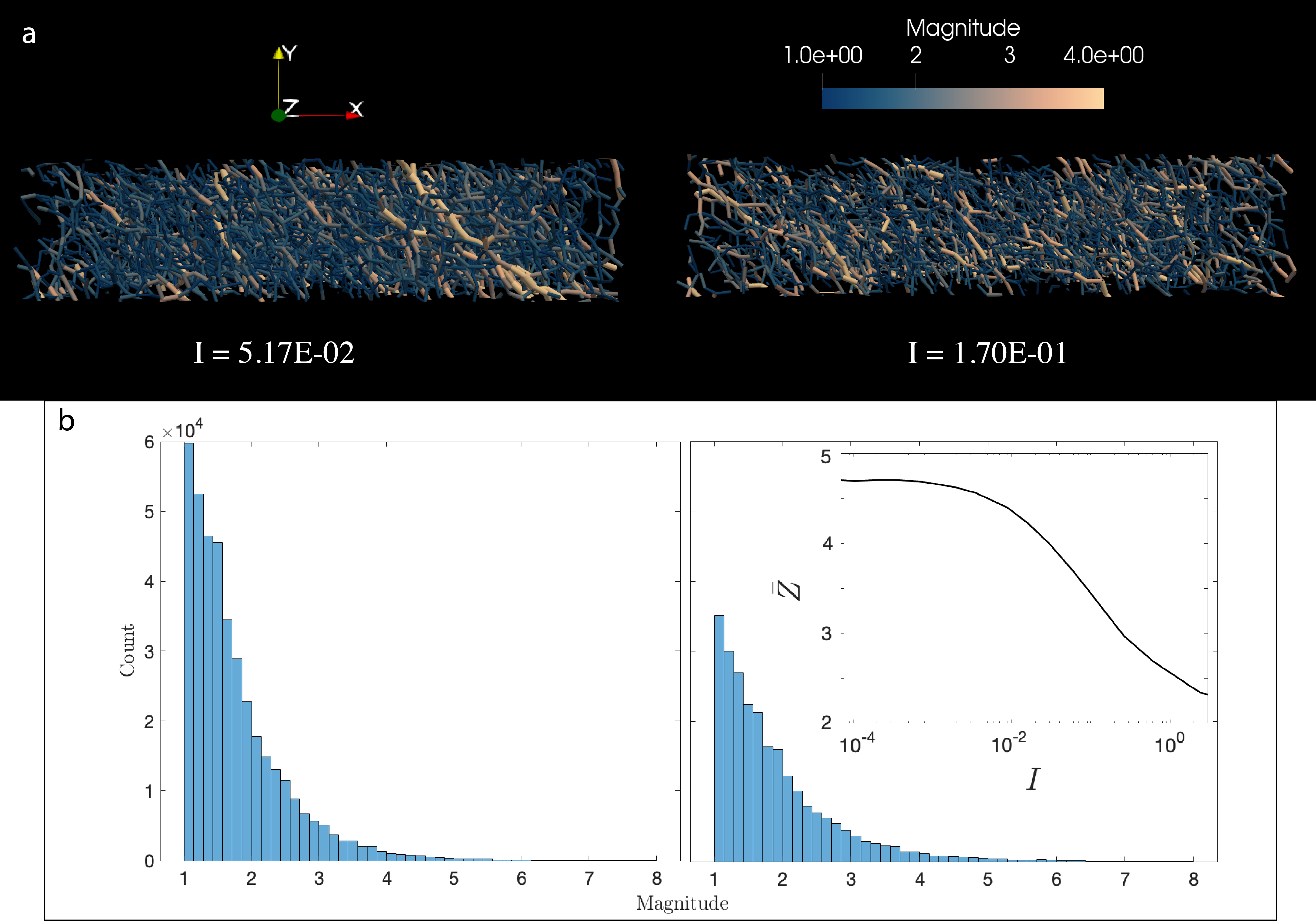}
\caption{(a) Visualization of force chains whose magnitude is non-dimensionalized by the average contact force for shear-cell simulations with $I\sim10^{-2}$ (left) and $I \sim 10^{-1}$ (right). Here, each contacting pair of particles is represented by a tube whose color and radius is scaled by the contact force normalized by the bed's weight. (b) Histograms of contact forces showing number of contacts binned via the non-dimensional magnitude. Inset: Time averaged coordination number $\bar{Z}$ as a function of $I$.} \label{FORCE_CHAINS}
\end{figure}

The plateau in $\Lambda$ that occurs at this glass-like transition (i.e. when $10^{-2}<I<10^{-1}$ and $\Phi(I)\rightarrow \Phi_c$; Figs. \ref{Lambda_I}d,e) is accompanied by a decrease in the rate of change in the ratio of the magnitude of the deviatoric kinetic and total stresses ($\sigma^{D}_{ki}$ and $\sigma^{D}$, respectively; Fig. \ref{SIGMA_SIGMA}). As $I>10^{-2}$, $|\sigma^D_{ki}|/|\sigma^D|$ becomes concave, suggesting a decrease in the rate at which kinetic stresses balance out shear-work with increasing $I$. This is reflected in the exponent to which the ratio scales with inertial number: when $I<10^{-2}$, the ratio scales linearly in log-log space as $I^{1.6}$, when $I>10^{-1}$ the ratio scales as $I^{1.4}$ (Fig. \ref{SIGMA_SIGMA}). In the state where $10^{-2}<I<10^{-1}$, the internal contact network is significantly altered, and the granular bed begins to dilate (Fig. \ref{RHEO}b). As this happens, the density and length of force chains decrease at the highest observable rate (Fig. \ref{FORCE_CHAINS}).  When $I<10^{-1}$, force chains with magnitudes greater than two times the granular bed's weight propagate through the entirety of the domain (Fig \ref{FORCE_CHAINS}a). When $I\sim10^{-1}$, contact forces are still correlated in space, but they are not as abundant (Fig \ref{FORCE_CHAINS}b). This behaviour can be further evidenced by the decrease in the average coordination number $\bar{Z}$, or the average number of neighbor-contacts a single particle has, as a function of $I$ (Fig. \ref{FORCE_CHAINS}b, inset). Thus, we suggest that shear work is consumed by the process of bed dilation, destroying aspects of the force-chain network during the `phase' transition from a contact dominated liquid-like granular phase, to a collisional gas-like flow (or, in other terms, as the flow transitions from the intermediate regime to the collisional regime). This process effectively inserts itself into the dissipative cascade of flow driving energy being converted into particle-scale energy fluctuations, which will ultimately be dissipated through irrecoverable deformation, frictional heating, and acoustic emissions \cite{Siman-Tov2021}. 

\begin{figure}[h!]%
\centering
\includegraphics[width = \textwidth]{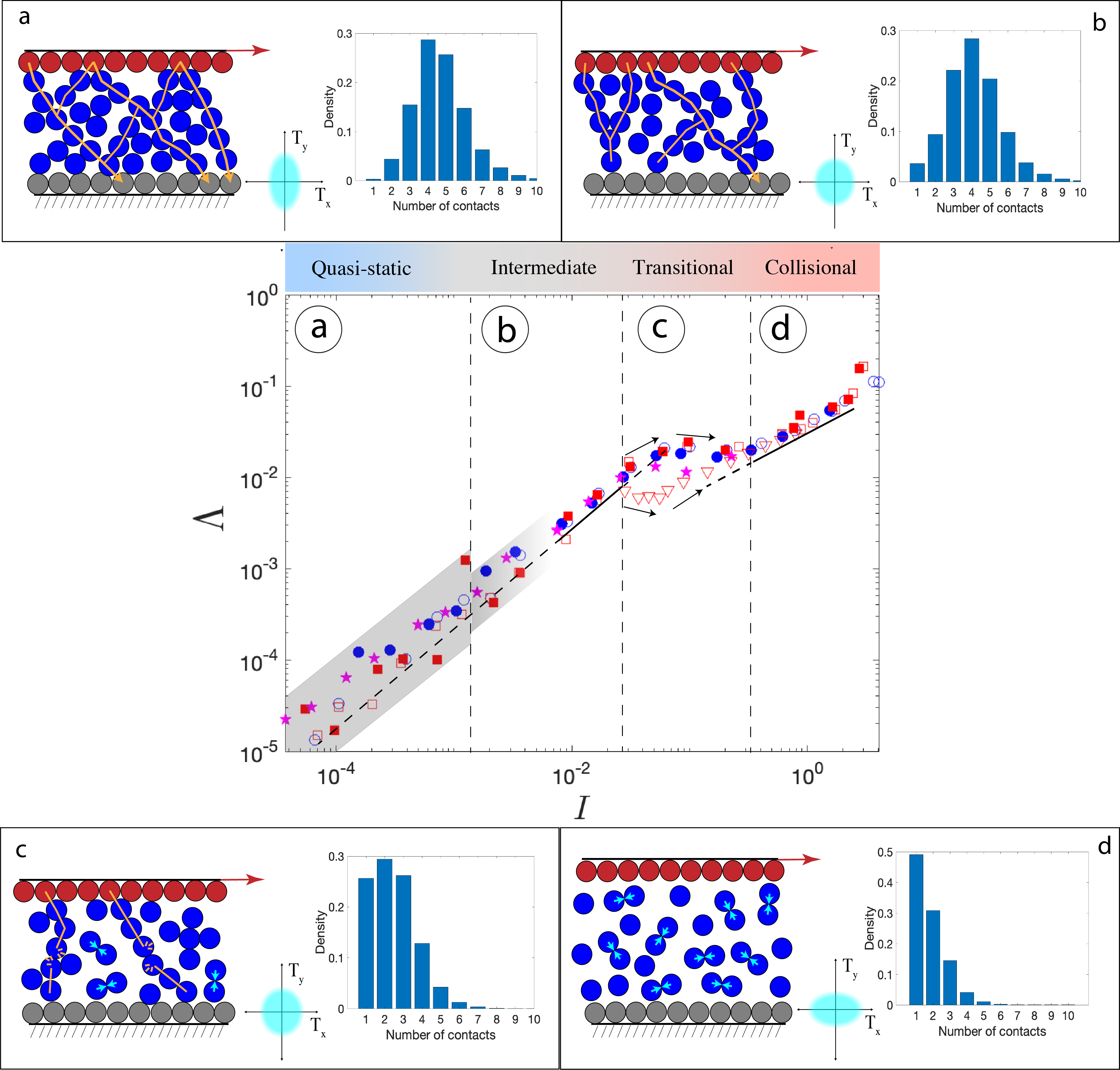}
\caption{Annotated Fig. \ref{Lambda_I}d, highlighting four regimes spanned by $\Lambda$, with bar charts showing the probability density of the number of particle contacts, and an axis representing the x and y components of granular temperature, where the cyan field qualitatively depicts the degree of anisotropy in that regime. Vertical dashed lines show regimes that are resolved by basal forces. (a) Force-chain dominated regime where $\Lambda$$\sim$ $I$. High magnitude domain spanning force chains cage particles; unsteady particle fluctuations, predominantly in the y-direction, disrupt the otherwise static architecture, weakening $\mu$. (b) Intermediate regime where linear log-log scaling persists and strengthens to the point of data collapse as bed deformation becomes continuous. (c) Transitional regime analogous to a latent heat-energy required to break force chains. (d) Collisional regime where binary particle collisions are the dominant mode of particle-particle and particle-wall interactions.}  \label{PHSSPC}
\end{figure}

Fig. \ref{PHSSPC} summarizes our data with an annotated $\Lambda(I)$ phase space (from Fig. \ref{Lambda_I}c). With this we can observe how changes in $\Lambda(I)$ are accompanied with transitions in and out of granular flow regimes. The first regime is defined when $I<10^{-3}$, aligning with the quasi-static regime that is described in the literature, and is characterized by many high magnitude, domain spanning, force chains (Fig. \ref{PHSSPC}a). Rheologically, this regime is concomitant with a decrease in apparent friction coefficient with increasing shear rates, both as a bed-averaged coarse-grained quantity and at the wall (Figs. \ref{RHEO}a and \ref{WALL_mu}a, respectively). This decrease in the ratio of stresses is also accompanied by a slight decrease in particle concentration (Fig. \ref{RHEO}b) and has been observed elsewhere, being attributed to endogenous vibrations coaxing particles to slip as they are near failure \cite{DeGiuli2017}. This is supported by the fact that granular temperature is dominant in the vertical direction (Fig. \ref{THETA}c), thus individual particle motion, `noise', works to dilate the bed of grains, decreasing the number of particles in static contact. This likely works in tandem with top plate motion stretching and rotating the long-lived force chains in the system, causing them to buckle and the bed to dilate \cite{Tordesillas2011}. We suggest that basal forces in this regime represent the diffusion of reordering events down to the base of the domain as shear-stresses generated by the top plate drive intermittent slip events \cite{daCruz2005}. $\Lambda(I)$ scaling in this regime shows higher degrees of dispersion about a general increasing log-linear trend, with an apparent particle size dependency. This shows that, in this framework, there remains particle scale effects that are due to the sensitivity of our results to spectral resolution: the mean frequency of the bed forcing signal is a function of $I$ (Fig. \ref{F_TIME}d), thus more energy is binned in lower frequencies with decreasing $I$. The lowest resolvable frequency is dictated by the length of the time-series $N$ and the sampling frequency $f_s$ ($\Delta f = f_s/N$), these do not change across simulations and particle size ($f_s = 100$ kHz and $N = 100000$). Thus, in this regime, particle size has a control on the frequency bands basal forcing power is relegated to, matching the fact that low inertial numbers signify when the time scale for particle rearrangements (function of particle size and confining pressure) dominates the kinematics of the system \cite{Midi2004}.

The second regime is characterized by $10^{-3}<I<10^{-2}$ (Fig. \ref{PHSSPC}b). Here, the bed of grains is characterized by continuous deformation and aligns with the aforementioned intermediate regime. In this regime, basal forcing is less controlled by particle scale parameters, but instead by the global kinematics of the system represented by the inertial number: linear log-log scaling persists, and is accompanied by the collapse of data points from disparate particle size simulations onto a single curve. Therefore, less scatter in the log-log linear $\Lambda(I)$ scaling marks the transition from quasi-static behaviour to continuous deformation of particles that remain in prolonged contact with one another. This is supported by the development of isotropic granular temperature, suggesting that energy fluctuations are controlled by those prolonged particle contacts (Campbell, 2006; Fig. \ref{THETA}c).

Increasing shear rates leads to a transitional regime where $10^{-2}<I<10^{-1}$ (Fig. \ref{PHSSPC}c). In past studies, this regime is included within the liquid-like intermediate regime. In this regime, the apparent coefficient of friction increases monotonically (Fig. \ref{RHEO}a), the particle concentration decreases rapidly (Fig. \ref{RHEO}b) and granular temperature remains isotropic in the absence of a viscous interstitial fluid (Fig. \ref{THETA}c). This aligns with previous definitions of the intermediate inertial regime \cite{Breard2020,Breard2024}. Nevertheless, as explained above, the scaling of basal force fluctuations with $I$ diverges from the log-log linear scaling observed elsewhere in the phase space. This transitional regime is reminiscent of latent heat of vaporization required to disrupt inter-molecular forces during the phase change from a liquid to a gas, thus leading to a dual existence interval where both domain spanning force chains and particle collisions contribute to forces recorded by the force plate. This interval exists until flow driving energy starts to become consumed by the dilation of the granular bed with increasing shear-work. Importantly, this regime contains the lowest $I$ values of the free surface inclined-slope flows, and $\Lambda$ measured in these simulations are displaced from the magnitude of values recorded in the shear-cell configurations, though general trends are maintained. 

The last regime spanned by our data reflects the inertial state transitioning into the gas-like collisional regime (Fig. \ref{PHSSPC}d). In the limit where $I>10^{-1}$, basal forces become noise-like, with a near flat spectral response across a wide band of frequencies (Fig \ref{F_TIME}a-c). Here, granular temperature becomes anisotropic in the stream-wise direction, suggesting the development of kinetic energy fluctuations through collisions and particle movement in the direction of shear-induced velocity gradients (Campbell, 2006; Fig. \ref{THETA}e-f). $\Lambda(I)$ scaling shows a collapse across particle sizes and configurations. This change in scaling behavior is mirrored by the scaling of $\Lambda$ with $\Phi$: as $\Phi < 0.57$ (exemplary of simulations with $I>10^{-1}$; Fig. \ref{RHEO}b) the $\Lambda(\Phi)$ scaling curves kinks, and the rate at which $\Lambda$ increases with decreasing solid concentration decreases. It is notable that in this regime, $\Lambda$ measured in the inclined-slope configurations joins the scaling curve of the shear-cell simulations. Thus, this transition suggests that as flow driving energy is used to dilate the flowing grains, preparation effects (or those effects that are configuration-dependent) occurring near the max packing volume fraction (e.g. \citeA{Josserand2000}) no longer play a strong role in fluctuating forces recorded at the base of the flow. This hints that structural memory is stored in elastic contacts in the form of the contact structures, and that this memory is erased as the flow enters into the collisional inertial gas-like regime. 

\FloatBarrier
\subsection{Towards real-scaled geophysical flows} \label{REASCALE}
\begin{figure}[h!]%
\centering
\includegraphics[width=\textwidth]{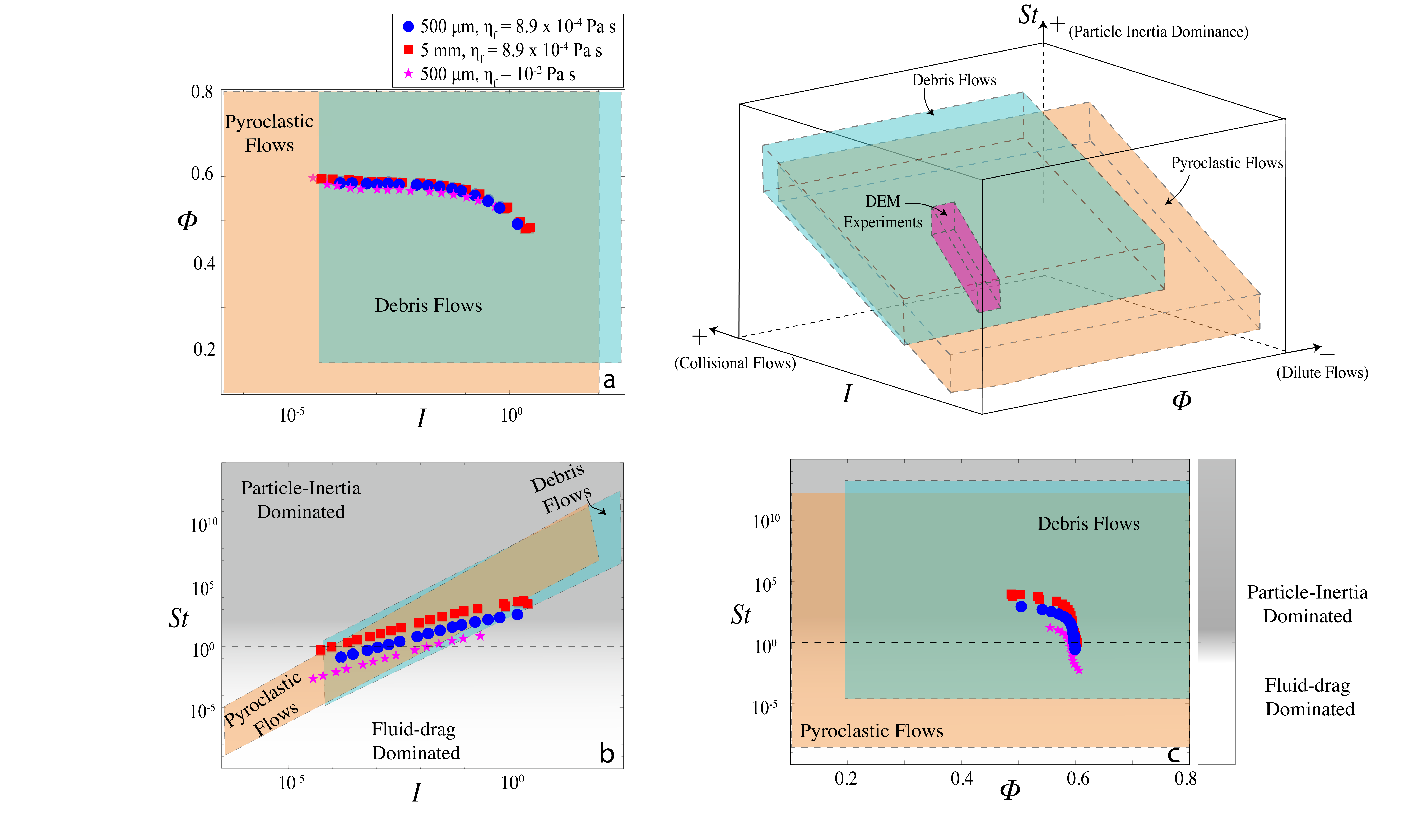}
\caption{Inertial number $I$, solid particle concentration $\Phi$, and Stokes number $St$ space. Shaded fields represent typical values for debris flows, pyroclastic flows, and discrete element method (DEM) simulations reported here. Sub-plots depict are taken from \citeA{Iverson1997,Iverson2001, Roche2012, Delannay2017, Breard2023}. (a-c) Display 2D phase spaces within this three-dimensional system.}  \label{REGIME}
\end{figure}

Geophysical mass flows can exhibit a wide range of inertial states, solid volume fractions, and can have interstitial fluid viscosities that span orders of magnitude. To visualize this, we sketch a three-dimensional space spanned by inertial number $I$, solid concentration $\Phi$, and Stokes number $St$, using typical values for velocity, flow height, particle density, fluid viscosity, solid concentration, and particle size to calculate the non-dimensional parameters (\citeA{Iverson1997, Iverson2001, Roche2012, Delannay2017, Breard2023}, Fig. \ref{REGIME}). Debris flows can span from the intermediate to collisional inertial regime, while having solid particle fractions that are on the order of 0.8-0.2. This reflects the fact that they can display behavior akin to highly frictional concentrated flows or very dilute muddy-water flows. Similarly, pyroclastic density currents can encompass a broad range of particle concentrations, spanning many orders of magnitude, from dusty-gas-like surges to highly concentrated flows that are exemplified by the dense basal undercurrents of pyroclastic density currents. In Fig. \ref{REGIME}, we have plotted dimensionless values for dense pyroclastic density currents known as pyroclastic flows, acknowledging that if we considered dilute pyroclastic density currents (known as surges), the shaded field representing these endmembers would extend far beyond what is sketched in the figure. 

Between debris flows and pyroclastic flows, characteristic particle size can span from $10^{-6}-1\ m$. Similarly, the viscosity of the interstitial fluid spans many orders of magnitude, with fluid compositions ranging from air (in the case of pyroclastic flows), with a viscosity of $\sim 10^{-5}\ Pa\ s$, to muddy water (in the case of debris flows), which can have a viscosity of $\sim 10^{-1}\ Pa\ s$ \cite{Iverson1997,Roche2012}. Thus, the Stokes number $St$, or the ratio of particle-inertial and fluid stresses, ranges from fluid stress dominated regimes (i.e., $St<<1$), to the end member where particle-inertia outpaces viscous fluid stresses (i.e., $St>>1$). Though the interstitial fluid viscosity of debris flows can be in excess of 100 times greater than water, typical values for debris flow height, characteristic particle size, and velocity suggest that under many conditions particle inertia overcomes viscous stresses. Conversely, the dilute, fines-enriched, endmembers of pyroclastic flows can exhibit Stokes numbers such that fluid stresses arising from relatively low fluid viscosities can overcome solid particle inertia. Nevertheless, there is much overlap in this three dimensional space between intermediate to dilute debris flows and the highly concentrated endmember of pyroclastic flows. With these disparate phenomena, we have plotted a field representing the values our simulations, showing that our data falls within the debris flow field, and within the field representing the most concentrated pyroclastic flows. With our discrete element simulations, we probe within this space conditions where fluid stresses are significant, while, as the prescribed top-plate velocity increases, conditions reach a point where particle-inertia outweighs fluid stresses, the inertial number increases, and the granular bed dilates, placing our data closer to conditions that align with highly inertial debris flows and concentrated pyroclastic flows. 

\begin{figure}[h!]%
\centering
\includegraphics[scale = 0.13]{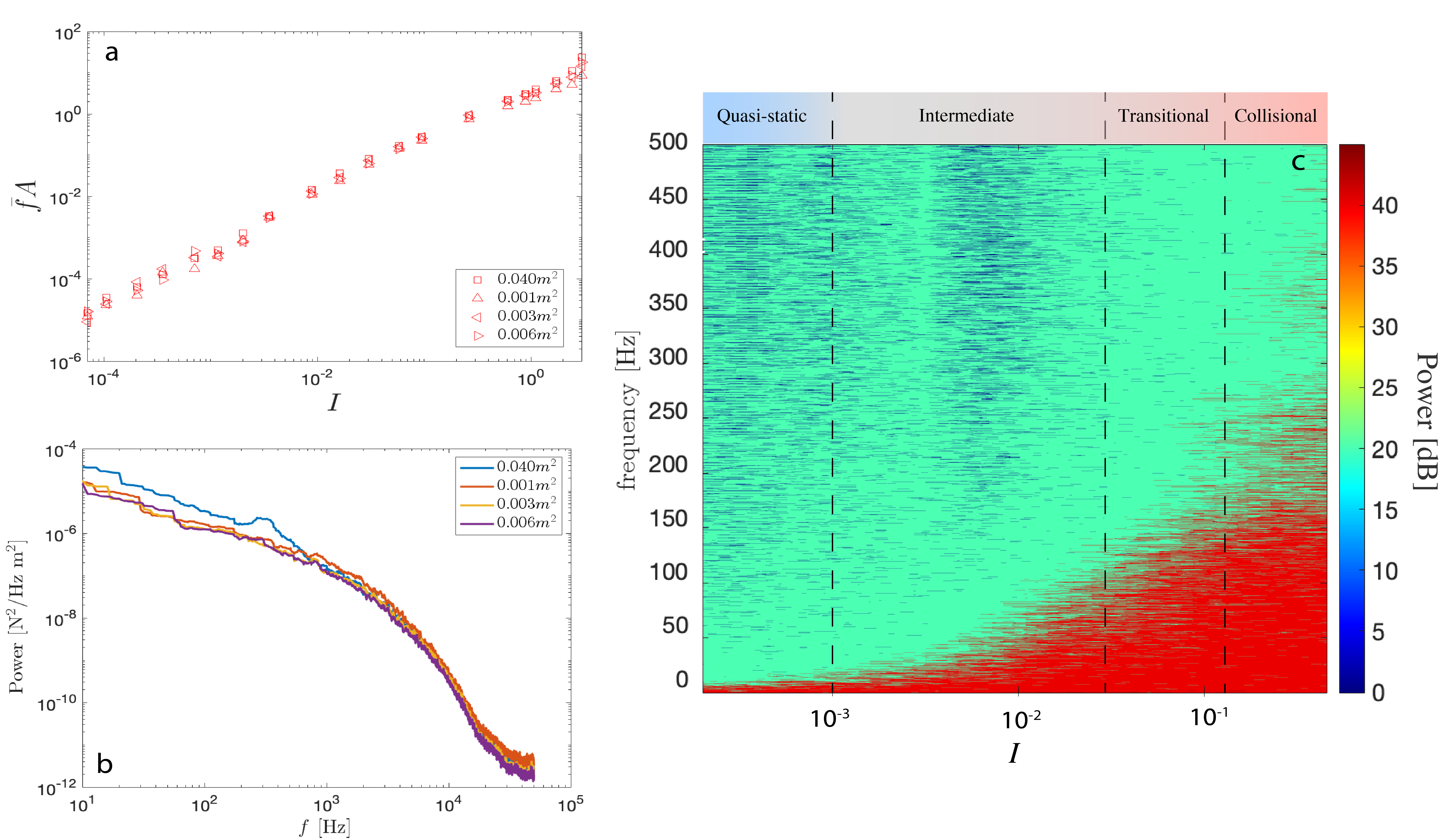}
\caption{(a) Mean frequency $\bar{f}$ of $5\ mm$ dry granular flow simulations, with variable simulated force plate sizes, scaled by the area of the force plate $A$, with areas equal to $0.001m^2$, $0.003m^2$, $0.006m^2$, and $0.04m^2$, as a function of inertial number $I$; (b) RMS power spectral density of basal forces derived from a dry granular flow simulation with $5\ mm$ particle and  $I$ $\sim$ $10^{-2}$, with curves showing the effect of changing the force plate size; (c) spectrogram of an idealized flow with inertial states such that $10^{-4}$ $<$ $I$ $<1$, with a footprint on the order of $1\ m^2$. Dashed lines indicate regimes resolved through bed force fluctuations (Fig. \ref{PHSSPC}).}  \label{SPGRM}
\end{figure}

Fig. \ref{REGIME} shows that our simulations achieve conditions that align with real-world flow conditions described by their typical inertial numbers, particle concentrations, and Stokes numbers. With the basal force data derived from these simulations, we can provide a first-order look into how basal-forcing power is binned in frequency space for an idealized flow on the scale of real-world flows. Here we examine a flow that develops from a pile of grains that exhibits solid-like properties, to a highly energetic continuous-flowing medium that is characterized by binary, gas-like, particle collisions. To do this, we will consider a hypothetical system with a footprint of $4\ m^2$. We will leverage the theoretical assumption that particle forces exerted onto the numerical force plate are uncorrelated. This assumption implies individual particle forces occur randomly, such that the shape of the forcing spectrum does not depend on the sum-total of individual particle forces, while the amplitude scales linearly with the square-root of the number of total forces \cite{Tsai2012}. Thus, we can scale our observed PSDs to the scale of geophysical flows via:
\begin{equation}
    P_{sc}(f) = \frac{A_p}{A_0}P(f)
\end{equation}
where $P_{sc}(f)$ is the scaled PSD, $A_p$ is the area of our hypothetical force plate ($4\ m^2$) and $A_0$ is the original area of the numerical force plate (we use the $5\ mm$ shear-cell simulations, therefore $A_0 =(40d)^2 = 0.04\ m^2$). We note that the assumption of uncorrelated forcings breaks down in flows that span inertial states from $10^{-4}<I<10^{-1}$, or where we observe long-range forces correlated in space (e.g., Fig. \ref{FORCE_CHAINS}). Nevertheless, Fig. \ref{SPGRM}a shows that the effect of decreasing a force plate's footprint has on the mean frequency $\bar{f}$ can be captured by scaling $\bar{f}$ by the area over which individual particle forces are summed (i.e., the force plate's area). Similarly, Fig. \ref{SPGRM}b shows the RMS of PSDs measured during a $5\ mm$ shear-cell simulation (where $I\approx10^{-2}$); when scaled by the numerical force plate area the general shape of the power spectrum does not change, though there are plate length-scale effect in frequencies less than $10^3\ Hz$ (these scale effects are absent when $I>10^{-1}$, Fig. S7). 

Using these scaled PSDs, we depict a spectrogram derived from our $5\ mm$  steady-state dry shear-cell simulations (Fig. \ref{SPGRM}). Note that this spectrogram is a stacking of 20 steady-state simulations, and represents an ideal; though the flow's inertial state appears to evolve in time, the data at every point in this figure represents basal forces from a steady developed flow. In this scenario, peak frequencies are well below $10\ Hz$ up to an inertial state of $I\sim10^{-2}$. This suggests that if all the basal forcing energy is radiated as seismic energy (i.e., simplifying Eq. \ref{conv} such that the detectable seismic signal does not depend on the green's function), the $5\ Hz$ nodal sesimometers used by, for example, \citeA{Allstadt2020} may not be able to detect signals generated by flows with inertial states less than $I<10^{-2}$. This has implications for monitoring the incipient stages of flow, when flow is defined by unsteady particle rearrangements ($10^{-4}<I<10^{-3}$), and the beginning phase of continuous motion when shear loads are no longer supported ($10^{-3}<I<10^{-2}$). At this stage, broadband seismometers may be needed to detect these events, if noise does not saturate the signal. Once the bed weakens and develops into a rapid liquid-like flow, where prolonged particle contacts are maintained while flow deformation is continuous ($I>10^{-2}$), the frequency content broadens and signals will likely be detectable by nodal deployments. 

\FloatBarrier
\section{Conclusions} 
Here we have recorded basal forces generated during dry and submerged granular flow simulations --- in plane-shear and inclined-slope flow configurations --- with the aim to correlate these signals to descriptions of the flows derived from coarse-graining procedures. Utilizing both rheologic and kinematic data from these simulations, we can describe four regimes, three of which have been been previously defined by granular physicists (e.g., \citeA{Midi2004}), and fourth regime that represents the transition from a liquid-like flow, characteristic of the intermediate inertial regime, to a gas-like collisional flow, defined by the collisional inertial regime (Fig. \ref{PHSSPC}). Therefore, our data suggests that macroscopic rheologic properties of granular flows are effectively encoded in forces exerted by the flow onto a simulated force plate. The four regimes are: (I) a solid-like regime that is characterized by the quasi-static inertial regime ($I<10^{-3}$), here there are long-lived force chain structures that span the entire flow domain, fluctuating basal forces are controlled by particle-scale parameters (i.e. size and hardness), granular temperature is anisotropic in the vertical axis which likely reflects endogenous noise generation as the stress-state (i.e., effective friction coefficient) decreases with increasing shear rates (velocity weakening); (II) an intermediate regime that corresponds to the intermediate liquid-like inertial regime, at this state the bed of particles deforms continuously, force chains build and break rapidly, the measure of total basal force fluctuations begins to collapse when expressed as a function of $I$, and granular temperature becomes isotropic reflecting particle contacts being the primary mode of granular temperature production; (III) a transitional regime that occurs in the canonical liquid-like inertial regime, showcasing a disruption in the conversion of flow driving energy into basal forcing as the flow transitions from the liquid-like to the collisional gas-like inertial state leading to a dual existence interval where boundary forcings are a function of both domain spanning force chains and individual particle impacts until the bed sufficiently dilates and the contact network is destroyed; and finally (IV) a regime characterized by the collisional inertial regime ($I>10^{-1}$). The inability for the flow to maintain prolonged and long range force chains, which ultimately leads to a data collapses in $\Lambda(I)$ space, suggests that regardless of flow configuration, the fluctuating basal forces are a function of inertial number. In summary, Fig. \ref{PHSSPC} suggests that basal tractions are well parameterized by kinematic quantities. 

We have shown that basal forces can encode detailed information about the dynamics of dry and submerged granular flows. Moreover, by examining the forces exerted on the force plate, we can delimit a transitional regime that is otherwise imperceptible in the typical $\mu(I)$ and $\Phi(I)$ rheology. Importantly, this effort provides a test bed that allows us to understand what rheological and kinematic data can potentially be extracted from these signals. Future work can utilize this understanding when analyzing forcing signals from well instrumented debris-flow channels and large-scale experiments (the reader is referred to \citeA{McCoy2013,Allstadt2020}). Further, the relationship between macroscopic rheological descriptions and basal forces can be leveraged into the development of sub-grid models of basal stresses. These models can be utilized in three-dimensional and depth-averaged models that can simulate flows on real-world scales, but whose resolution and derivation cannot accurately depict granular stresses that lead to substrate erosion and seismicity. Moreover, our data provides a clue as to how inverted basal tractions from seismic signals generated by debris flows may be used to elucidate dynamical and rheologic information about a given flow, providing another tool for probing these complex and dangerous phenomena. 

  \begin{notation}
  \notation{Bold indicates vector quantity}
  
  \notation{$\tilde{A}$}
  Discrete amplitude spectrum
  \notation{$A_0$}
  Original area of the numerical force plate
  \notation{$A_p$}
  New area of the numerical force plate
  \notation{$C_d$}
  Drag coefficient
  \notation{$d$}
  Particle diameter
  \notation{$dt_s$}
  DEM solver solid time-step
  \notation{$dt_f$}
  DEM solver fluid time-step
  \notation{$e_n$}
  Normal restitution coefficient
  \notation{$\mathbf{F}_c$} 
  Contact force
  \notation{$\mathbf{F}_d$} 
  Drag force
  \notation{$\mathbf{F}_d^{i \in k}$}
  Drag force on the $i^{th}$ particle in the $k^{th}$ computational cell
  \notation{$F^n_{k_n}$} 
  Normal elastic force
  \notation{$F_\eta^n$} 
  Normal dissipative force
  \notation{$F(t)$} 
  Total basal force time series
  \notation{$F_x(t)$} 
  Total basal forces in the x-direction
  \notation{$F_y(t)$}
  Total basal forces in the y-direction
  \notation{$\tilde{F}(f)$}
  Fourier transform of total basal forces
  \notation{$f_c$}
  Total basal forcing corner frequency
  \notation{$f_{ny}$}
  Total basal forcing Nyquist frequency
  \notation{$f_s$}
  Sampling frequency
  \notation{$\bar{f}$}
  Mean total basal forcing frequency 
  \notation{$\Delta f$}
  Frequency bin width
  \notation{$\mathbf{g}$}
  Gravitational acceleration
  \notation{$H$}
  Granular flow thickness
  \notation{$I$}
  Inertial number
  \notation{$I_v$}
  Viscous number
  \notation{$I_m$}
  Modified inertial number
  \notation{$\mathbf{I}_{f}$}
  Total interphase momentum exchange term
  \notation{$\mathbf{I}_{f}^k$}
  Interphase momentum exchange term in the $k^{th}$ cell
  \notation{$k_n$}
  Normal particle elastic stiffness coefficient
  \notation{$K$}
  Particle-Cell weight kernel
  \notation{$m_i$} 
   Mass of $i^{th}$ particle
   \notation{$m_{ef}$}
   Effective mass of contacting particles
   \notation{$N_s^k$}
   Number of solid phase particles in the $k^{th}$ cell
   \notation{$N$}
   Number of discrete points in time series
   \notation{$N_{Sav}$}
   Savage number
   \notation{$N_{Bag}$}
   Bagnold number
   \notation{$P_f$}
   Fluid pressure
   \notation{$P_p$}
   Prescribed top plate confining pressure
   \notation{$P_s$}
   Solid pressure
   \notation{$P(f)$} 
   Power spectral density of basal forces
   \notation{$P_{sc}(f)$}
   Scaled power spectrogral density 
   \notation{$Re^k$}
   Particle Reynolds number in the $k^{th}$ cell
   \notation{$\overline{\overline{S}}_f$}
   Fluid phase stress tensor
   \notation{$St$}
   Stokes number
   \notation{$t$}
   Time
   \notation{$t_n^{col}$}
   Timescale of normal particle collisions
   \notation{$T$}
   Isotropic granular temperature 
   \notation{$T_x$}
   Granular temperature in the x-direction
   \notation{$T_y$}
   Granular temperature in the y-direction
   \notation{$T_i^w$}
   Granular temperature at the force plate wall in the $i^{th}$ direction
   \notation{$\mathbf{u}_f$}
   Fluid velocity 
   \notation{$\mathbf{V}_i$}
   Velocity of the $i^{th}$ particle
   \notation{$V_n^i$}
   Velocity of the $i^{th}$ particle in the normal direction
   \notation{$V_n^j$}
   Velocity of the $j^{th}$ particle in the normal direction
   \notation{$\mathbf{V}_f$}
   Interpolated mean fluid velocity
   \notation{$V_p$}
   Prescribed top plate velocity
   \notation{$v_i^w$}
   Force plate wall-averaged instantaneous slip velocity in the $i^{th}$ direction
   \notation{$V_i^w$}
   Time averaged force plate wall-averaged slip velocity in the $i^{th}$ direction
   \notation{$\mathbf{X}_{i\in k}$} 
   Location of $i^{th}$ particle in the $k^{th}$ cell
   \notation{$\mathbf{x}_k$} 
   Location of grid node centered on the $k^{th}$ cell
   \notation{$\bar{Z}$}
   Time-averaged coordination number
   \notation{$\beta_s^k$}
   Interphase exchange coefficient in the $k^{th}$ cell
   \notation{$\dot{\gamma}$}
   Shear rate
   \notation{$\epsilon_f$}
   Fluid volume fraction
   \notation{$\epsilon_{s}$}
   Solid volume fraction 
   \notation{$\eta_n$}
   Normal particle viscous dampening coeff.
   \notation{$\eta_f$}
   Fluid phase viscosity
   \notation{$\theta$}
   Slope
   \notation{$\theta_s$}
   Angle of repose
   \notation{$\Theta$}
   Non-dimensional granular temperature
   \notation{$\Theta_w$}
   Non-dimensional granular temperature at the force plate wall
   \notation{$\Lambda$}
   Non-dimensional basal force fluctuations
   \notation{$\mu$}
   Effective friction coeff. 
   \notation{$\mu_p$}
   Particle friction coeff. 
   \notation{$\mu_w$}
   Effective friction coeff. at the force plate wall
   \notation{$\nu_k$}
   Volume of the $k^{th}$ cell
   \notation{$\nu_s$}
   Volume of solid phase particles
   \notation{$\rho_f$}
   Fluid density 
   \notation{$\rho_s$}
   Particle density 
   \notation{$\sigma_{xy}$}
   Shear stress
   \notation{$\sigma^D$}
   Total deviatoric stress 
   \notation{$\sigma^D_c$}
   Deviatoric contact stress 
   \notation{$\sigma^{ki}_c$}
   Deviatoric kinetic stress
   \notation{$\tau_p$}
   Particle deformation timescale
   \notation{$\tau_{micro}$}
   Microscopic particle rearrangement timescale
   \notation{$\tau_{macro}$}
   Macroscopic flow deformation timescale
   \notation{$\Phi$}
   Coarse grained solids concentration
   \notation{$\Phi_c$}
   Critical solids concentration
  \end{notation}

\section{Open Research}
The DEM-CFD code used to reproduce this data is publicly available at \sloppy{https://mfix.netl.doe.gov} and can be downloaded after registering with the Department of Energy website here: \sloppy{https://mfix.netl.doe.gov/register/}. Readers can find more information, and movie visualizations of example simulations in the supporting information. Example post-processing scripts and simulation input files can be found in \citeA{Zrelak2024}.

\acknowledgments
Financial support to P.Z. and J.D. provided by NSF EAR grants 1926025 and 1949219. P.Z. was further supported by NASA award 80NSSC20K1773. E.C.P.B. was supported by UKRI with the NERC-IRF (NE/V014242/1). We thank University of Oregon's High Performance Computing cluster Talapas for providing support and the infrastructure to produce the data. The authors would also like to thank the reviewers for their feedback and thoughtful comments that strengthened this effort.  

\raggedright
\nocite{duPont2003}
\bibliography{Zrelak_et_al}
\end{document}